\documentclass[fleqn,usenatbib]{mnras}

\usepackage{newtxtext,newtxmath}
\usepackage{amsmath}

\usepackage[T1]{fontenc}

\DeclareRobustCommand{\VAN}[3]{#2}
\let\VANthebibliography\thebibliography
\def\thebibliography{\DeclareRobustCommand{\VAN}[3]{##3}\VANthebibliography}

\usepackage{graphicx}	
\usepackage{amsmath}	

\newcommand{\msun}{M$_{\odot}$}	

\newcommand{\zta}{$\zeta$}
\newcommand{\fy}{$\phi$}
\newcommand{\mzout}{$\dot{M}_{Z}$}

\newcommand{\kmps}{km s$^{-1}$}

\newcommand{\aref}[1]{\hyperref[#1]{Appendix~\ref{#1}}}

\title[QED I]{\textsc{Quokka}-based Understanding of Outflows Derived from Extensive, Repeated, Accurate, Thorough, Demanding, Expensive, Memory-consuming, Ongoing Numerical Simulations of Transport, Removal, Accretion, Nucleosynthesis, Deposition, and Uplifting of Metals (QUOD ERAT DEMONSTRANDUM, i.e, QED). I. Metal loading, phase structure, and convergence testing for Solar neighbourhood conditions}

\author[Vijayan et al.]{
Aditi Vijayan$^{1,2}$\thanks{E-mail:aditi.vijayan@anu.edu.au},
Mark R. Krumholz$^{1,2}$,
Benjamin D. Wibking$^{3}$
\\
$^{1}$Research School of Astronomy and Astrophysics, Australian National University, Canberra ACT 2601, Australia\\
$^{2}$ARC Centre of Excellence for Astronomy in Three Dimensions (ASTRO-3D), Canberra ACT 2601, Australia\\
$^{3}$Department of Physics and Astronomy, Michigan State University, 567 Wilson Road, East Lansing, MI 48824, USA\\
}

\date{Accepted XXX. Received YYY; in original form ZZZ}

\pubyear{2015}

\begin{document}
\label{firstpage}
\pagerange{\pageref{firstpage}--\pageref{lastpage}}
\maketitle

\begin{abstract}
Multiphase galactic outflows, generated by supernova feedback, are likely to be more metal-rich than the interstellar media from which they are driven due to incomplete mixing between supernova ejecta and the ambient ISM. This enrichment is important for shaping galactic metallicities and metallicity gradients, but measuring it quantitatively from simulations requires resolution high enough to resolve mass, momentum and energy exchanges between the different phases of the outflows. In this context, we present simulations of outflows, driven by SN feedback, conducted using \textsc{Quokka}, a new GPU-optimised AMR radiation-hydrodynamics code. This code allows us to reach combinations of resolution, simulation volume, and simulation duration larger than those that have previously been possible, and to resolve all gas phases from cold neutral medium, $T \sim 100$ K, to hot ionised gas, $T \gtrsim 10^7$ K. In this, a first of a series of papers exploring generation and evolution of multiphase outflows from a wide range of galactic environments and star formation rates, we quantify the extent of selective metal loading in Solar neighbourhood-like environments. We explain the selective metal loading we find as a result of the transport of metals within and between phases, a phenomenon we can study owing to the parsec-scale resolution that our simulations achieve. We also quantify the sensitivity of metal loading studies to numerical resolution, and present convergence criteria for future studies.
\end{abstract}




\section{Introduction}

Almost all elements heavier than helium, i.e. metals, are manufactured in stars. From their point of origin, metals travel far and wide to populate not only the interstellar medium (ISM) within and the circumgalatic medium (CGM) around galaxies, but also the inter-galactic medium (IGM) pervading the cosmos. Because the distribution of metals is driven by several physical processes, such as outflows from stellar feedback, inflows of metal-poor gas from the IGM, and mixing of different gas phases, metals act as tracers of these phenomenon. By following the life-cycle of metals as they traverse a galaxy, we can hope to understand the complex process of galaxy evolution.

An important observational result that links metal abundances to large-scale galaxy evolution is the mass-metallicity relation (MZR; \citealt{Tremonti+2004}), a relatively tight correlation between gas phase metallicities and galaxy stellar masses that extends over several orders of magnitude in stellar mass. The metallicity, expressed in terms of the oxygen abundance of the gas phase, increases sharply with stellar mass at low galactic masses (albeit with a great deal of scatter) and flattens out at larger masses. In dwarf galaxies one possible explanation for this correlation is the preferential removal of metals in supernova-driven outflows \citep[e.g.][]{Peeples&Shankar2011, Zahid14a, Christensen18a, Forbes19a}. \citet{Peeples&Shankar2011}
quantify the selective enrichment of galactic outflows relative to the mean metallicity of the ISM in terms of the ``metal expulsion efficiency'' or ``metal loading factor'', $\zeta$, which mathematically is ratio of the metallicity of the outflowing gas to that of the gas present in the ISM. Models with $\zeta \gg 1$ have the advantage that they do not require the extreme mass loading factors (often $\gtrsim 100$) that are required to explain the mass-metallicity relation in dwarfs in more conventional models where $\zeta = 1$, i.e., where outflow and ISM metallicities are assumed to be the same \citep[e.g.][]{Finlator&Dave2008, Dave12a, Lilly13a}. Moreover, radially-resolved models with $\zeta = 1$ tend to produce relatively steep metallicity gradients in dwarf galaxies, in contradiction to the observed flatness of dwarf gradients (the mass metallicity gradient relation, MZGR; e.g., \citealt{Belfiore17a, Mingozzi20a, Poetrodjojo21a}); by contrast, models including selective metal loading naturally reproduce the MZGR in dwarfs \citep{Sharda21a, Sharda21b, Sharda23a}.

While these analyses hint at the importance of selective metal loading for galactic properties, both direct observational and numerical explorations have been limited. With regard to the former, though observations of outflows in star-forming galaxies are ubiquitous \citep[e.g.][and references therein]{Veilleux20a}, to date there are only limited observational constraints on the metal-loading (or mass-loading) factor, and then only for a limited range of outflowing gas phase. 
For the warm ionised phase, \citet{Chisholm+2018} find for a sample of seven galaxies over a wide range in stellar mass that the outflow metallicity can exceed the ISM metallicity by as much as a factor of $50$ in dwarf galaxies, though this number might be smaller for larger galaxies. They also find that the bulk of the metals in the outflows arise from entrained ISM, rather than direct SN ejecta. However, metal loading factors derived from absorption studies are fraught with uncertainties related to parameters such as geometry and ionization correction. \citet{Cameron+2021} circumvent uncertainties related to ionization by directly estimating the electron temperature from auroral lines in the winds of Mrk $1486$, an edge-on disc galaxy with biconical outflows \citep{Duval+2016}. They find that outflows along the minor axis of the galaxy are enriched with metals compared to the ISM of the disc. It is worthwhile to note that though both \cite{Cameron+2021} and \cite{Chisholm+2018} conclude that the outflows in Mrk $1486$ are metal enriched, their estimates of the degree of enrichment differ by a factor of two, likely indicative of the level of observational uncertainty.

There is also significant evidence for selective metal enrichment in studies of the faster, hotter components of outflowing gas.
\citet{Martin02a} and \citet{Stevens03a} find that X-ray emitting gas in the winds of the dwarf starbursts NGC $1569$ and M82, respectively, has a higher $\alpha$ to iron ratio than those galaxies' interstellar media, strongly suggesting the presence of incompletely-mixed type II supernova ejecta in the wind.
Recently, \cite{Lopez+2020} analysed detailed \textit{Chandra} spectra of M82 to estimate the metal content of the different temperature phases of its multiphase outflowing gas. Consistent with the conclusions drawn from the aforementioned UV studies of the warm ionised phase, and with the earlier X-ray work, they find that the outflows show gradients in metallicity, which differ by temperature. The relatively cooler warm-hot ($\gtrsim 10^6$ K) phase, traced by O and Ne lines, retains near-Solar abundance similar to that in the disc, while the hot ($>10^7$ K) phase, traced by Si and S, is enriched relative to the disc by up to a factor of $\sim 3.5$, although these results are at least somewhat sensitive to the fitting procedure \citep{Ranalli+2008, Konami+2011}. 

Differential metal loading of the phases of the outflowing gas is an idea supported by theoretical works as well. \cite{Melioli+2013} study 3D simulation of galactic winds in a starburst system. They follow chemical evolution of the outflows and find that the metallicity of the lower density (higher temperature) phase may be nearly $4.5$ times larger than that of the higher density (lower temperature) phase. A recent study by \cite{Emerick+2018} that 
tracks detailed chemical evolution of ejecta from not only type II SN but also AGB stars in an isolated dwarf galaxy also conclusively shows that metal enrichment preferentially takes place in the hotter gas phase. \citet{Andersson23a} and \citet{Rey23a} reach similar conclusions in simulations reaching higher resolutions -- the former reaches a maximum resolution of $1.5$ pc, but only in the densest regions of the galactic midplane, while the latter has a peak resolution of $18$ pc but maintains this resolution well into the outflow. These findings are in line with the prediction from the seminal paper by \cite{MaclowAndrea99} that smaller galaxies lose nearly all, if not all, the metals from SN ejection. Long-term simulation of such outflows, explored by \cite{Melioli+2015} (see also \citealt{Fragile+2004, Rodriguez+2011}), further underscore that the metal-enriched SN ejecta escapes into IGM and may lead to its enrichment \citep{Aguirre+2001}.

While isolated galaxy simulations offer deep insights into the overall budget of metals between a galaxy and its surroundings, they lack the spatial resolution required to accurately follow the exchange of metals amongst the phases of the outflowing gas -- a physical process that occurs at pc or even smaller scales \citep[e.g.,][]{Gentry19a}. Capturing this exchange requires resolutions at this level not just in the galactic plane, but far into the outflow \citep{Vijayan+20}, and is required for reliable simulation-observation comparison, since observations are generally sensitive only to particular gas phases. In this respect, tall-box simulations emulating a smaller patch of a galactic disc, but offering $\sim$pc scale resolution, are better poised to study metal fluxes \citep{Kim&Ostriker2017, Li&Bryan2017}.

Even the best-resolved of the tall-box simulations carried out to date, however, have limited resolution or limited volume. \cite{Creasy+15} study metal enrichment of galactic winds in an aggregate sense through a suite of simulations that explore the parameter space of gas surface density and gas fraction. Their results indicate that the metal loading of winds emanating from a larger galaxy is weaker even though the absolute metallicity of the winds is higher. However, while they achieve $2$ pc resolution, their simulation volume extends to only $\pm 500$ pc around the galactic midplane, meaning that they cannot study the phase structure of the outflow except very near the plane. \citet{Li&Bryan2017} use a suite of tall-box simulations to understand the relationships between mass, energy, and metal loading factors and the underlying gas properties. For their fiducial run, which replicates Solar neighbourhood conditions, they achieve a maximum resolution of $2$ pc only within $500$ pc of the midplane, and the resolution worsens to $8$ pc beyond $|z|>1$ kpc. In order to successfully capture these between different temperature phases, simulations with uniformly high resolution are an imperative.

The simulations of metal loading of outflows offering the highest combination of volume and resolution published to date are those of \citet{Kim+20} and \citet{Schneider20a}. The former carry out simulations of metals in galactic winds with a uniform resolution of $256^2 \times 1792$ with spatial scales of $2$ or $4$ pc per cell, corresponding to simulation regions that extend to either $\pm 1.8$ or $\pm 3.6$ kpc around the midplane. The latter use $2048^2 \times 4096$ cells in a $10\times 10\times 20$ kpc domain, achieving 5 pc resolution, but with a $10^4$ K temperature floor so the simulations omit neutral material. However, the literature to date lacks a rigorous convergence study identifying convergence criteria and demonstrating converged measurements of metal loading of galactic outflows; the most through study to date is that of \citet{Schneider20a}, who find that quantities such the phase structure remain unconverged at their highest resolution of $5$ pc. Thus it is unclear if the results from any published simulations are converged.

In this paper we overcome the challenge of high computation costs of high resolution simulations by using the state-of-the-art GPU-based adpative mesh refinement (AMR) code, \textsc{Quokka}, which is significantly faster than CPU-based codes \citep{QuokkaMethods}. We use this code to examine metal loading of winds in a star-forming galaxy using uniformly-high resolution tall-box simulations, reaching combinations of numbers of cells and resolution comparable to or better than the best published to date. We distill the simulation results to express the metal-loading of galactic outflows using two metal-loading factors that track different routes of metal enrichment of outflows, and we demonstrate that our estimates for these quantities are converged in resolution. This paper is the first in a series using \textsc{Quokka} to study the transport of metals in outflows, the \textsc{Quokka}-based Understanding of Outflows Derived from Extensive, Repeated, Accurate, Thorough, Demanding, Expensive, Memory-consuming, Ongoing Numerical Simulations of Transport, Removal, Accretion, Nucleosynthesis, Deposition, and Uplifting of Metals (QUOD ERAT DEMONSTRANDUM, or QED for short). This first paper focuses on quantifying metal loading and phase structure in uniformly-resolved (i.e., non-AMR) simulations of Solar neighbourhood conditions, and on testing for convergence in these simulations. Subsequent papers will explore the use of adaptivity in wind simulations, and will extend the study to other galactic environments.

The remainder of this paper is organised as follows. In \autoref{sec:methods} we describe our numerical methods. In \autoref{sec:results}, we present our primary results regarding both the physics of metal loading and the numerics of achieving converged measurements of it.

\section{Methods}
\label{sec:methods}

\subsection{Simulation setup and initial conditions}

We conduct 3D HD simulations using \textsc{Quokka} \citep{QuokkaMethods}, an adaptive mesh refinement (AMR) radiation-hydrodynamic code optimised for GPUs. For the purposes of this paper we do not include radiative transfer, only radiative cooling. For its hydrodynamic step, \textsc{Quokka} solves the Euler equations of compressible gas dynamics using a method of lines formulation with an RK2 update that is second-order accurate in time and space. We also include gravitational forces provided by a static potential representing the stars and dark matter in a galactic disc; in the present work we do not include gas self-gravity.

All our simulations take place in a $1 \times 1 \times8$ kpc$^{3}$ domain, with the longest dimension along the $z-$axis and the galactic plane centred at $z=0$. We use periodic boundary conditions in the $x$ and $y$ directions, and first-order extrapolation boundary conditions in the $z$ direction. For this first study we do not use the AMR capability of \textsc{Quokka} in order to ensure that our simulations have uniformly high resolution throughout the wind and to make testing for convergence straightforward; we will extend this study to AMR simulations in a future work. For this paper, our resolution is uniform throughout the volume, and our cell sizes range from $\Delta x$ from $32$ to $2$ pc, corresponding to resolutions from $32^2 \times 256$ to $512^2 \times 4096$ cells. We summarise the properties of the simulations in \autoref{tab:params}, and for convenience from this point on we refer to the simulations as FG$N$, where $N$ is the resolution in pc and FG indicates that we are using fixed (i.e., non-adaptive) grids. For comparison, our highest resolution case, FG2, contains $\approx 10\times$ as many cells as the simulations of \citet{Kim+20}. It contains a factor of $4$ fewer than the simulations of \citet{Schneider20a}, but has $2.5\times$ higher resolution and runs for a substantially longer time ($\approx 120$ Myr as opposed to $70$ Myr), allowing it to reach statistical steady-state. 

\begin{table*}
\begin{center}
\begin{tabular}{|l|c|c|c|c|c|}
\hline
\hline
Name & $\Delta x$ (pc) & $N_{x}N_{y}N_{z}$  & $\phi$ &  $\zeta ({Z_{\rm bg}=0})$ &  $\zeta ({Z_{\rm bg}=Z_\mathrm{O,\odot}})$ \\ 
(1)  & (2) & (3) & (4) & (5) & (6) \\
\hline
\hline
FG$32$ & $32$   & $32^2 \times 256$   & $0.54^{0.54}_{0.53}$ & $14^{15}_{13}$ & $1.4^{1.4}_{1.3}$\\
\\
FG$16$ & $16$   & $64^2 \times 512$   & $0.74^{0.78}_{0.70}$ & $12^{14}_{10}$ & $1.2^{1.3}_{1.2}$\\
\\
FG$8$ & $8$   & $128^2 \times 1024$   & $0.95^{1.0}_{0.90}$ & $16^{18}_{14}$ & $1.3^{1.3}_{1.2}$\\
\\
FG$4$ & $4$   & $256^2 \times 2048$   & $0.96^{1.0}_{0.86}$ & $41^{46}_{37}$ & $1.6^{1.6}_{1.5}$\\
\\
FG$2$ & $2$  & $512^2 \times  4096$  & $0.83^{0.93}_{0.75}$ & $26^{28}_{24}$ & $1.3^{1.4}_{1.3}$\\
\hline

\end{tabular}
\caption{
\label{tab:params}
Summary of runs. (1) Name of the run; (2) Base resolution of the grid in pc; (3) Number of cells in each direction on the base grid; (4) Steady state value of the metal loading factor \fy~(\autoref{eqn:phi}); (5) \& (6) Steady state value of the metal loading factor \zta\ for $Z_{\rm bg} =0$ or $Z_\mathrm{O,\odot}$ (\autoref{eqn:zeta}). For columns (4)-(6), the central value we report is median over times from $100-116$ Myr (after steady-state has been established), and the super- and subscripts indicate the temporal $84-$th and $16-$the percentile, respectively. 
}
\end{center}
\end{table*}

The initial density and pressure profiles are adapted from the Solar Neighbourhood model of TIGRESS simulations \citep{Kim&Ostriker2017}. In this model, the initial density profile follows a double exponential representing a two-phase medium, 
\begin{equation}
    \rho(z) = \rho_{1,0} \exp\left(-\frac{\Phi_{\rm ext}(z)}{\sigma_1^2}\right) 
    + \rho_{2,0} \exp\left(-\frac{\Phi_{\rm ext}(z)}{\sigma_2^2}\right),
\end{equation}
where $\sigma_{1,2}$ are the sound speed of the two phases, here set to $7$ \kmps\ and $70$ \kmps, respectively. The midplane densities are $\rho_{1,0} =2.85 m_{\rm H}$ cm$^{-3}$ and $\rho_{2,0}=10^{-5}\rho_{0,1}$. The external gravitational potential, $\Phi_{\rm ext}$, is set by the dark matter halo potential and a stellar disc. The dark matter potential is adapted from \citet{Kuijken89a}, and the total potential from the dark matter halo and the stellar disc (reproduced from \citealt{Kim&Ostriker2017}) is,
\begin{equation}
    \Phi_{\rm ext} =
    2\pi G \Sigma_* z_* \left[ \left( 1 + \frac{z^2}{z_*^2} \right)^{1/2}-1\right]
    + 2\pi G \rho_{\rm dm} R_0^2 \rm{ln}\left(1+ \frac{z^2}{R_0^2}\right).
\end{equation}
Here, $\Sigma_*= 42$ M$_{\odot}$ pc$^{-2}$, $z_*= 245$ pc, $\rho_{\rm dm}=6.4 \times 10^{-3}$ M$_{\odot}$ pc$^{-3}$ and $R_0$ is the Galactocentric radius of our simulation box, which we set to be $8$ kpc. With these choices of $\rho_{1,0}$, $\rho_{2,0}$, $\sigma_{1,2}$, and $\Phi_\mathrm{ext}$, the initial gas surface density, $\Sigma_{\rm gas}$, is $13$ M$_{\odot}$ pc$^{-2}$

As noted above, while we do not include radiative transfer, we do include radiative heating and cooling. We implement these using a custom cooling source term that is similar, but not identical, to that used by the \textsc{Grackle} library \citep{Grackle}. We cannot use the \textsc{Grackle} code itself because \textsc{Grackle} does not run on GPUs. Instead, we adopt the tabulated primordial and metal line heating and cooling tables that are included with \textsc{Grackle}, and we re-implement the tabular interpolation routines, as well as the terms for photoelectric heating and Compton cooling (which are not included in the tables themselves) with a temperature floor of $100$ K. However, unlike \textsc{Grackle}, we do not include an X-ray heating term. We then integrate the cooling function in each cell in an operator-split manner using \textsc{Quokka}'s adaptive RK integrator that runs on GPUs. The model used in \textsc{Quokka} and \textsc{Grackle} includes photoelectric heating at a rate that matches that in the Solar neighbourhood, and produces an atomic medium with well-defined warm and cold phases whose properties are in reasonable agreement with observations; see Appendix A of \citet{Wibking23a} for a detailed discussion of the cooling model and a comparison between it and other commonly-used approaches.

\subsection{Implementing supernova feedback}\label{sec:SNFeedback}

We implement SN feedback by injecting thermal energy equivalent to a single SN event, $10^{51}$ erg, into a single cell in the simulation domain. We also add a fixed density of passive scalar $\Delta Z_\mathrm{SN}/V_\mathrm{cell}$ to the same cell, where $V_\mathrm{cell}$ is the cell volume, representing metal injection due to SNe; note that the value of $\Delta Z_\mathrm{SN}$ is arbitrary, as we will discuss in \autoref{sec:quantify_met_loading}. The total number of such feedback events is determined by the star formation rate corresponding to the initial gas surface density. We use a star formation rate density of $6\times 10^{-3}$ \msun\ kpc$^{-2}$ yr$^{-1}$ which is the SFR used by \cite{Li&Bryan2017} for Solar neighbourhood conditions. For a \cite{Chabrier2001} IMF, the corresponding surface rate density of SN events is $\Sigma_\mathrm{SN} = 6\times 10^{-5}$ kpc$^{-2}$ yr$^{-1}$, and the total SN rate in our simulation box is therefore $\Gamma_\mathrm{SN} = 6\times 10^{-5}$ yr$^{-1}$.

SNe are distributed randomly in the $x-y$ plane and in the $z-$direction their distribution follows a Gaussian with a width of  $150$ pc. Thus the SN probability per unit volume per unit time is\begin{equation}
    \mathcal{P}(z) = \frac{d^2 P}{dt dV} = \overline{N} e^{-\frac{z^2}{h^2}}\,,
\end{equation}
where $h=150$ pc and $\overline{N}=\Sigma_\mathrm{SN} h^{-1} \pi^{-1/2}$, so that $\int \mathcal{P} \,dz = \Sigma_\mathrm{SN}$. For this probability density, the expected number of SNe per time step $dt$ in a volume $V_\mathrm{cell}$ is
\begin{equation}
    \langle N \rangle = \Sigma_{\rm SN} h^{-1} \pi^{-1/2} V_\mathrm{cell}\, dt\,.
\end{equation}
In practice we implement SN as follows: for each time step of size $dt$ on the coarsest AMR level, we compute the number of SNe that will occur during that time step by drawing from a Poisson distribution with expectation value $\Gamma_\mathrm{SN}\, dt$. For each SN that is to go off during this time step, we determine the position by drawing from a uniform distribution in the $x$ and $y$ directions and from Gaussian distribution with width $h$ in the $z$ direction. 
We then add thermal energy and passive scalar to the cell enclosing the coordinates for that SN. 

The SN events occurring in the initial few Myr of the simulation run create a hot phase. Because this is a volume-filling phase, subsequent SN events are more likely to occur in a low density region of the disc, ensuring that the snowplow radius of the supernova remnant is larger than the resolution of the simulation \citep{Forbes+16}. We emphasise that we do not provide any sub-grid treatment of SN feedback, e.g., injecting radial momentum or kinetic instead of thermal energy. Such models are unnecessary at the resolutions we reach, and are undesirable because they are necessarily resolution-dependent, which can make it impossible to test for or achieve convergence.

\subsection{Quantifying metal loading}\label{sec:quantify_met_loading}

In this section our goal is to explain how we quantify the degree of metal loading in our simulations.

\subsubsection{Computing the metal abundance}

Before we can compute metal loading, we must first calculate the metallicity in each cell of the simulation. This will be determined by two factors: the initial metal abundance present at the start of the simulation, to which we refer as $Z_\mathrm{bg}$, and the metal mass added per SN, to which we refer as $\Delta M_{Z,\mathrm{SN}}$. Note that, thanks to the setup of our simulation, we are free to choose these factors \textit{ex post facto} -- that is, since SNe inject only passive scalar and not mass, and since we know both the total mass density and the passive scalar density in each cell, we are free to compute the metallicity in the cell as an arbitrary linear combination of these two densities after the simulation has already run. Specifically, we write the metal density in every cell as
\begin{equation}
    \label{eqn:rho_Z}
    \rho_Z = Z_\mathrm{bg} \rho + \frac{\Delta M_{Z,\mathrm{SN}}}{\Delta Z_\mathrm{SN}} \rho_s,
\end{equation}
where $\rho$ is the total mass density and $\rho_s$ is the passive scalar density. Note when a SN goes off in a cell the scalar density $\rho_s$ in that cell increases by $\Delta Z_\mathrm{SN}/V_\mathrm{cell}$, so $\rho_Z$ increases by $\Delta M_{Z,\mathrm{SN}}/V_\mathrm{cell}$, as it should.

In practice we choose values of $Z_\mathrm{bg}$ and $\Delta M_{Z, \mathrm{SN}}$ appropriate for oxygen. The oxygen output of a single type II SN is $\approx 1$ M$_\odot$ \citep{Nomoto+2013}, and we therefore adopt $\Delta M_{Z,\mathrm{SN}} = 1$ M$_\odot$. We consider a range of values of $Z_\mathrm{bg}$, which we parameterise in terms of the Solar oxygen abundance: $Z_\mathrm{bg} = (Z/Z_\odot) Z_{\mathrm{O},\odot}$, where $Z_{\mathrm{O},\odot} = 8.6\times 10^{-3}$ \citep{Asplund09}. Below we consider values of $Z/Z_\odot$ from $0$ to $2$.

\subsubsection{Definitions of metal loading factors}

\begin{figure}
\begin{center}
    


	\includegraphics[width=\columnwidth]{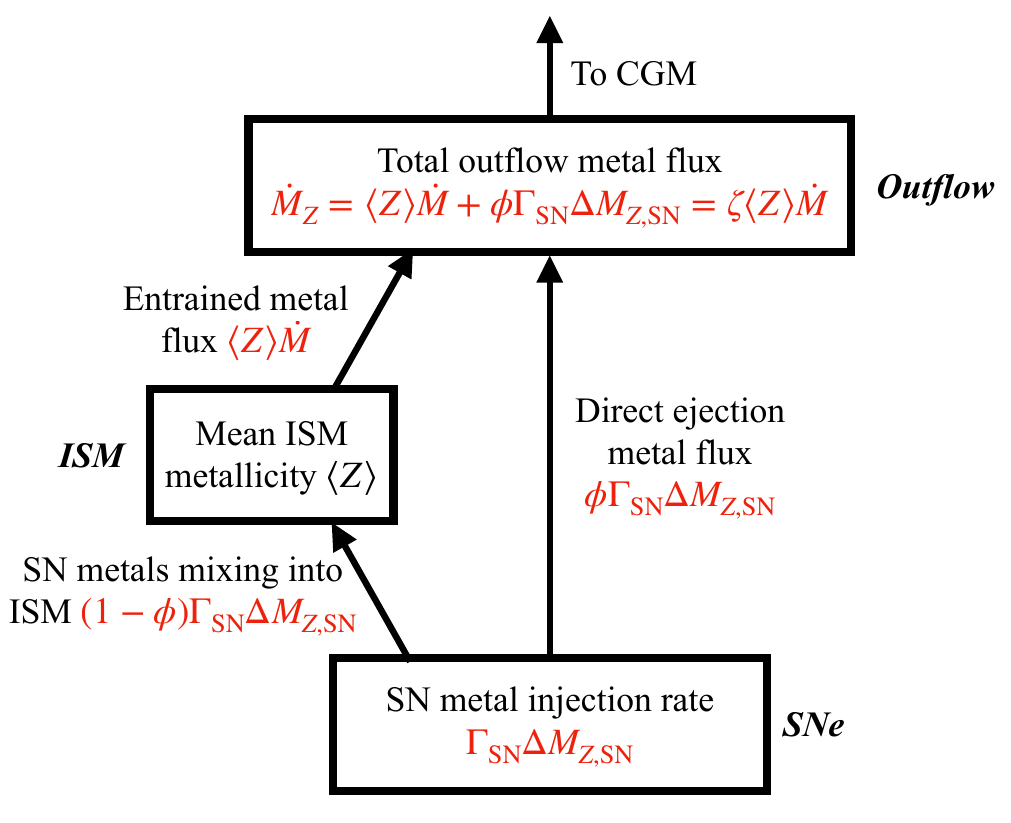}
    \caption{A schematic depicting the meaning of our two metal loading factors, $\phi$ (\autoref{eqn:phi}) and $\zeta$ (\autoref{eqn:zeta}), and their relationship to the metal injection rate and outflow metal flux.}
    \label{fig:zeta_phi_schematic}
\end{center}
\end{figure}

Now that we have defined the metallicity $\rho_Z$ in each cell, we are in a position to define the key quantity that we wish to extract from our simulations, the ``metal loading factor'', generally taken to be the ratio between the metal flux and the star formation rate in the galaxy (see for example Equation 7 of \cite{Li&Bryan2017}, and also \cite{Kim+20}, for a definition based on fluxes). However, such a definition assumes that all the metal outflows are solely sourced from the SN activity. A galaxy goes through multiple episodes of star formation over its lifetime, each of which adds to the ambient metallicity of the ISM gas in which future SN events will occur. These events in turn will entrain some of the previously enriched ISM into outflows. Under such circumstances, the metal outflows comprise contributions from not only direct SN ejecta but also the entrainment of enhanced-metallicity ISM, and we wish to define metal loading factors that can track these two channels independently.

\autoref{fig:zeta_phi_schematic} provides a schematic illustration of the picture that motivates our definitions: SNe inject metals at a rate $\Gamma_{\rm SN} \Delta M_{\rm {Z,SN}}$ into the region close to the midplane of the galaxy which lies at the bottom of the figure. These metals can end up in either of two boxes labeled ``ISM'' or ``Outflow''. Metals that end up in the ISM box are mostly retained by the galaxy and enhance its overall abundance. However, as gaseous outflows are established in the galaxy, some of these retained metals may be entrained with the mass outflows, contributing an amount $\langle Z\rangle \dot{M}$ to the total metal flux, where $\langle Z\rangle$ is the mean metallicity of the ISM gas being entrained, and $\dot{M}$ is the mass outflow rate for entrained ISM gas. By contrast, metals produced by SNe that are ejected directly into outflows do not mix with ISM and instead escape the disc -- this process is represented by the ``Outflows'' box. The net metal outflow rate $\dot{M}_{Z}$ includes both the entrained ISM and direct outflow components. By contrast, the total mass flux is overwhelmingly dominated by the entrained component, since direct SN ejecta carry very little mass; i.e., the total mass outflow rate is simply $\dot{M}_\mathrm{out}$, the same as the outflow rate for entrained ISM. 

To account for the different contributions to the metal outflows, we introduce two different factors to quantify metal loading, viz, $\zeta$ and $\phi$, which we define as follows. First, we define the net metal outflow rate through a plane of fixed height $z$ at time $t$ as
\begin{equation}
\label{eqn:mzdot}
    \dot{M}_{Z} (z,t)= \iint \rho_Z  v_z \, dx \, dy,
\end{equation}
where $v_z$ is the $z-$velocity of the gas and $\rho_{Z}$ is the metal density. We note here that the net outflow rate as given here includes both inflowing and outflowing gas. However, once steady-state has been achieved most of the gas is outflowing. As a result, the results do not change substantially if we instead consider only outgoing gas. Similarly, the mass outflow rate through the surface at height $z$ is
\begin{equation}\label{eqn:mdot}
    \dot{M} (z,t)= \iint \rho  v_z \,dx \, dy\,,
\end{equation}
where $\rho$ is the total mass density.

Since our system is symmetric about $z=0$, at least statistically, in practice we will always use $[\dot{M}_Z(z,t) + \dot{M}_Z(-z,t)]$ in place of simply $\dot{M}_Z(z,t)$, and similarly for $\dot{M}$, i.e., we should always understand that when we write $\dot{M}_{Z}$ or $\dot{M}$ at a given $z$ and $t$, the quantity we intend is actually the sum of the fluxes through the $+z$ and $-z$ surfaces.

We then separately estimate the contribution to the metal flux from entrained ISM as $\langle Z \rangle \dot{M}$, where $\langle Z \rangle $ is the average metallicity of material bounded between $-z$ and $+z$, i.e.,
\begin{equation}\label{eqn:avg_z}
    \langle Z \rangle \equiv \frac{\int_{-z}^z \iint \rho_{Z} \, dx\, dy\, dz }{\int_{-z}^{z} \iint \rho \,dx\,dy\,dz}
\end{equation}
Using these equations, we can define our metal loading factor $\zeta$ as the ratio of the metal outflow rate to the outflow rate that would be expected if the outflows consisted purely of entrained ISM, i.e.,
\begin{equation}
    \label{eqn:zeta}
    \zeta = \frac{\dot{M}_{Z}}{\langle {Z} \rangle \dot{M}}\,.
\end{equation}
For example, $\zeta=2$ corresponds to a situation where the metal flux in the outflow is twice what would be expected if the outflow consisted purely of entrained ISM. This quantity is analogous to the $\zeta$ factor defined by \citet{Peeples&Shankar2011}. By contrast, the factor $\phi$ quantifies the fraction of SN metal output that is directly added to outflow without ever mixing with the ISM. We define this quantity as 
\begin{equation}
    \label{eqn:phi}
    \phi = \frac{\dot{M}_{ {Z}} - \langle Z\rangle \dot{M}}{\Gamma_{\rm SN} \Delta M_{\rm {{Z},SN}}} = \left(\frac{\zeta-1}{\zeta}\right) \frac{\dot{M}_{{Z}}}{\Gamma_{\rm SN} \Delta M_{\rm {{Z},SN}}}\,.
\end{equation}
Here the numerator can be interpreted as the metal outflow rate \textit{subtracting off} the contribution from entrained ISM, while the denominator is the total metal injection rate by SNe. Thus for example a factor $\phi = 0.3$ corresponds to a case where $30\%$ of the metals injected by SNe are never mixed into the ISM, and are instead lost promptly; this quantity is analogous to the SN yield reduction factor introduced by \citet{Sharda21a}.

\section{Results}
\label{sec:results}


\subsection{Qualitative simulation outcomes}

\begin{figure*}
	\includegraphics[width=\textwidth]{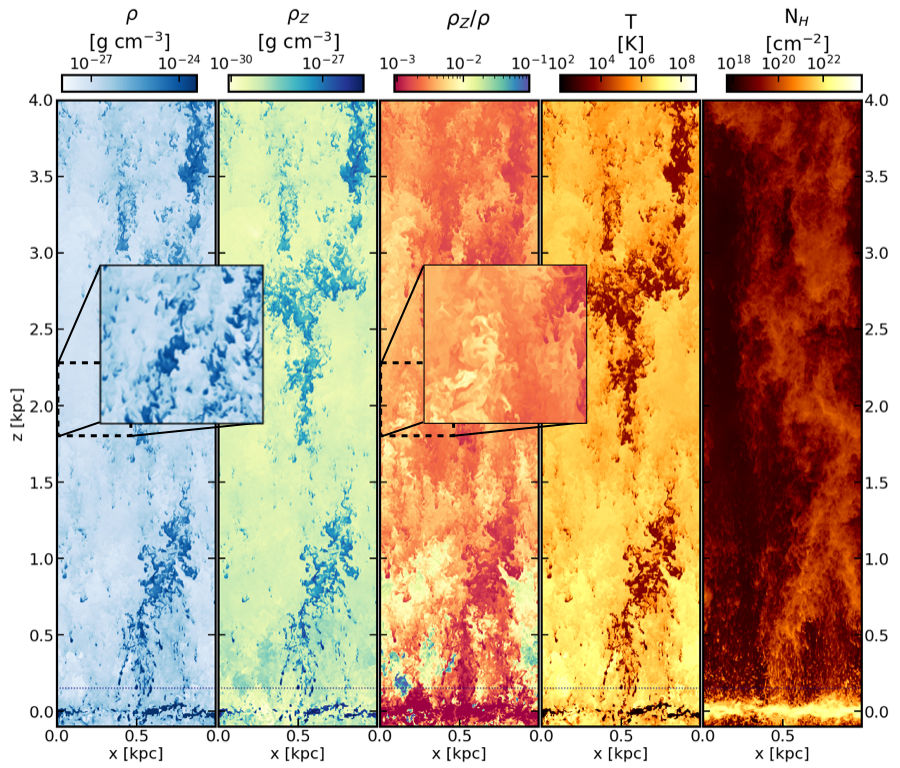}
    \caption{Slices of gas density, metal density, metal abundance, and temperature in run FG$2$ at time $t=115$ Myr from left to right, respectively. The right column shows the column density along the $y$ axis. The horizontal line indicates the initial scale height of the disc. For reasons of space we show only the top half of the simulation domain, but remind readers that the domain extends to $-4$ kpc below the plane as well. Inset panels zoom in on example regions a high resolution.}
    \label{fig:slice_plot}
\end{figure*}

\begin{figure}
	\includegraphics[width=\columnwidth]{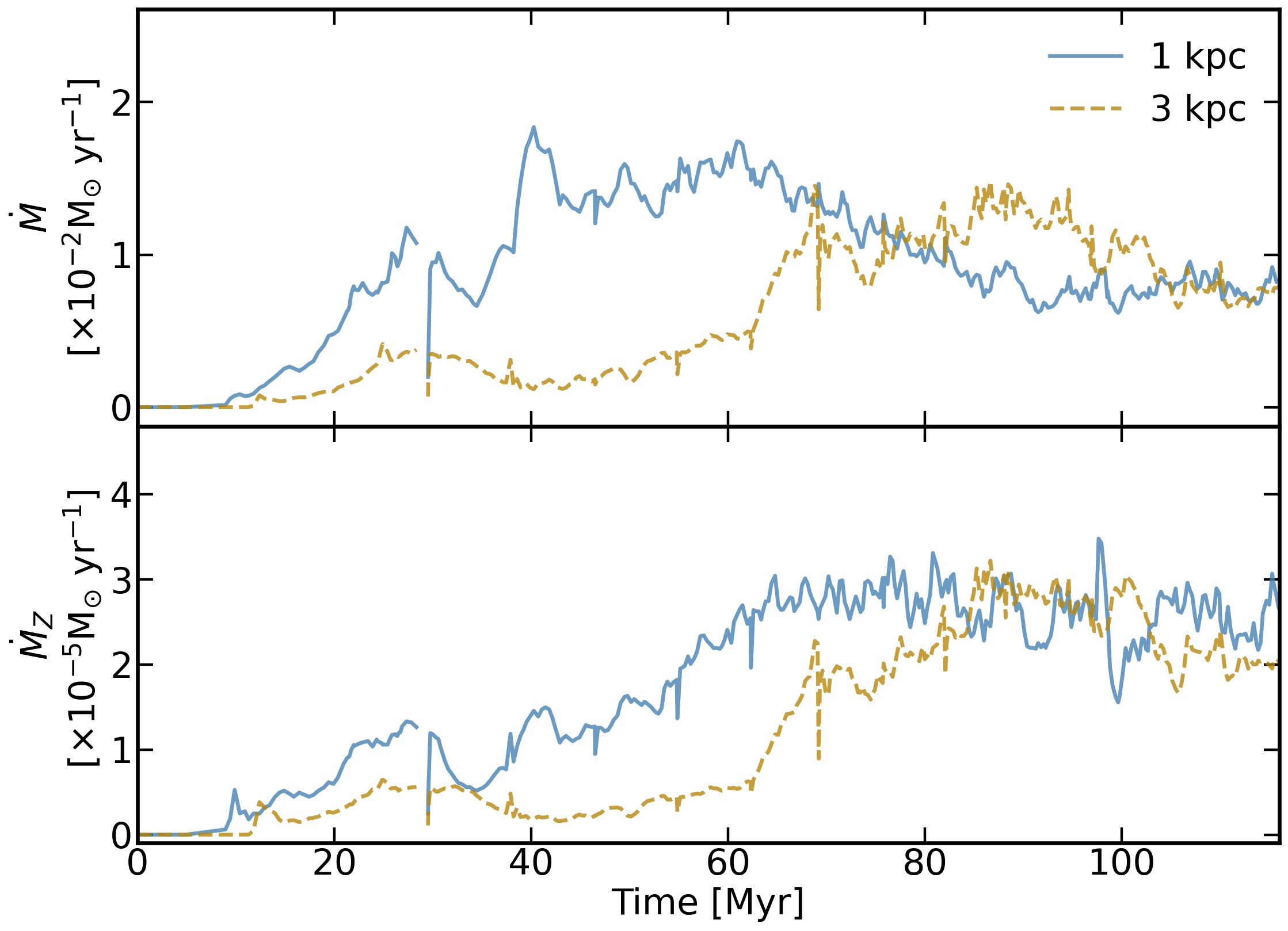}
    \caption{Mass (top) and metal (bottom) outflow rates through surfaces at $z=1,3$ kpc, computed from \autoref{eqn:mzdot} and \autoref{eqn:mdot} for FG$2$. The figure shows that, after an initial transient, the outflow rates settle down to near-steady-state values at times $\gtrsim 100$ Myr.}
    \label{fig:outflow_rates}
\end{figure}

We begin by describing the qualitative outcome of our simulations in order to orient the reader for the quantitative analysis that follows. For this purpose we make use of run FG$2$ evaluated with $Z_\mathrm{bg} = 0$, though we note that the qualitative behaviour is the same in all runs, and that for phenomena where the value of $Z_\mathrm{bg}$ matters we will show multiple sample values. As SN feedback begins in the system, hot bubbles develop around the injection sites. Within a few Myr, the individual bubbles expand and break out of the disc. SN feedback produces a volume-filling hot gas and subsequent SN explode into this medium. Disc-wide outflows are set up in the galaxy and the initially stratified medium turns multiphase. 

\autoref{fig:slice_plot} shows slices of gas density ($\rho$), metal density ($\rho_Z$), metallicity ($\rho_Z/\rho$), and temperature and the column density along the $y$ axis at a time when steady outflows have been set up in the galaxy. In the outflowing gas, we identify the warm, dense gas likely lifted from the disc. The hotter parts of the outflows, arising from direct injection of SN, are metal-enriched while the cooler parts are comparatively metal-poor. As these different phases propagate out of the disc, they mix and produce regions of metal-poor warm gas surrounded by hot metal-enriched gas; these features are seen particularly clearly in the inset panels, which zoom in on some of the cool clouds. From the column density we can make out the disc comprising cool, dense gas.


We plot the mass and metal outflow rates, as computed from \autoref{eqn:mzdot} and \autoref{eqn:mdot}, through the $|z|=1,3$ kpc surfaces as a function of time in \autoref{fig:outflow_rates}. To reduce noise the outflow rates are averaged over a thickness of $5$ cells both above and below the surfaces. Both mass and metal fluxes rise initially as individual superbubbles break out and outflows escape from the disc, with the rise occurring first at $|z| = 1$ kpc and then later at $|z|=3$ kpc. After $\sim 100$ Myr of evolution sustained outflows of mass and metals are set up in the entirety of the simulation domain. The system achieves a near steady-state around this time as subsequent outflow rates fluctuate only at a factor of $\lesssim 2$ level, although there is a slow secular decrease in the mass outflow rate due to the loss of gas mass from the simulation domain through the boundaries at $z=\pm 4$ kpc. The mass and metal fluxes through the $1$ kpc and $3$ kpc surfaces are very similar, as expected given that we are plotting net mass fluxes. However, even if we plot outward-going only mass fluxes, the results are not substantially different, indicating that most of the material that reaches a height of $1$ kpc also reaches $3$ kpc. The total steady-state mass outflow rate, $\dot{M} \approx 0.5 - 1 \times 10^{-2}$ M$_\odot$ yr$^{-1}$, is comparable to the star formation rate that corresponds to our chosen supernova rate, $\dot{M}_* = 6 \times 10^{-2}$ M$_\odot$ yr$^{-1}$. Thus the overall mass loading factor in our simulation is $\approx 1$.

\subsection{Bulk loading factors}

\begin{figure}
	\includegraphics[width=\columnwidth]{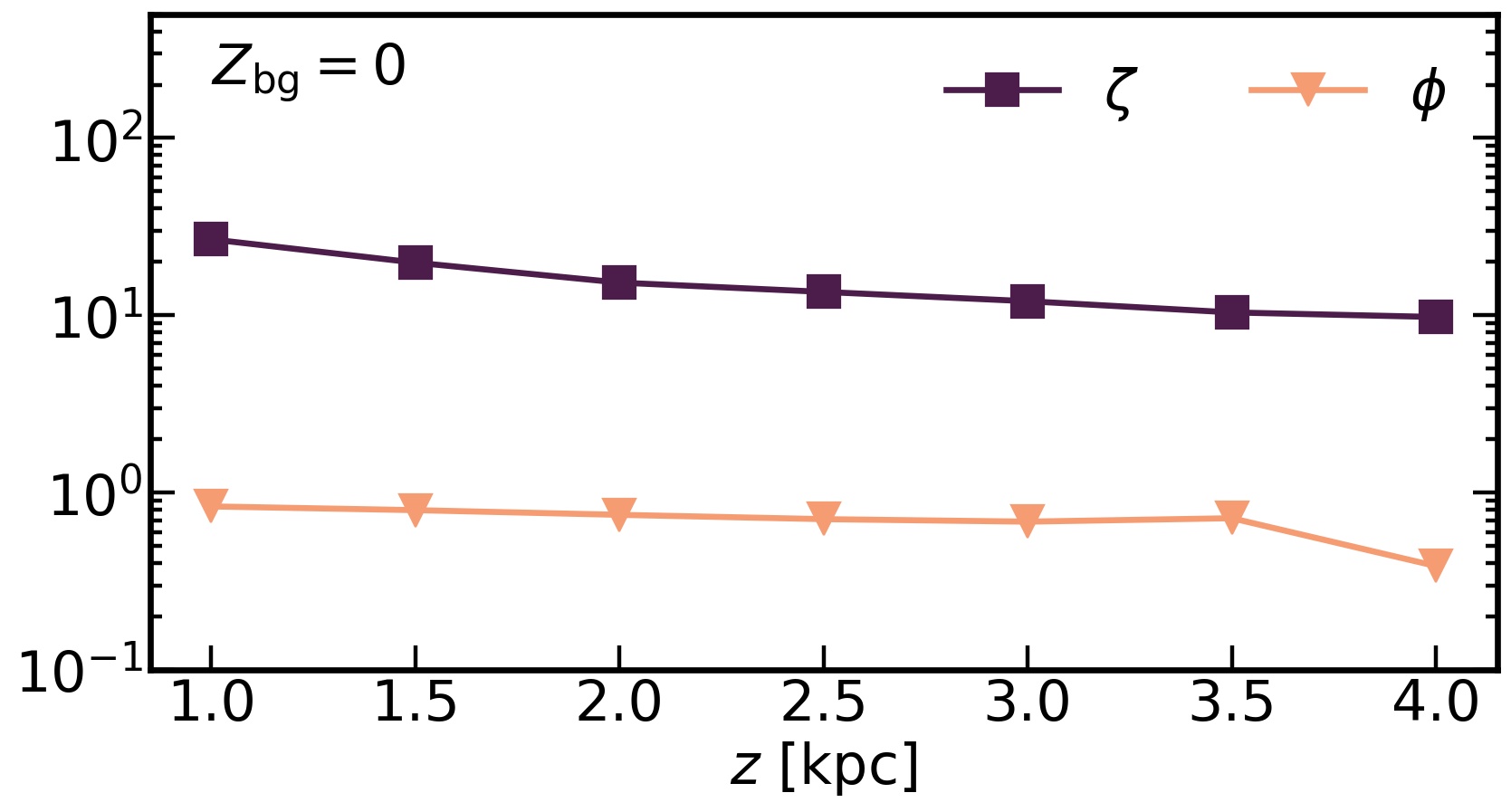}
    \caption{Metal loading factors \zta\ (\autoref{eqn:zeta}) and \fy\ (\autoref{eqn:phi}) for a pristine background, $Z_\mathrm{bg} = 0$. $\zeta \gg 1$ indicates that metals outflows are dominated by fresh SN ejecta rather than entrained ISM gas, while \fy\ values  close to unity indicate that most of the metals added to the galaxy by feedback are lost to outflows. 
    }
    \label{fig:zeta_phi_height}
\end{figure}

\begin{figure}
\includegraphics[width=\columnwidth]{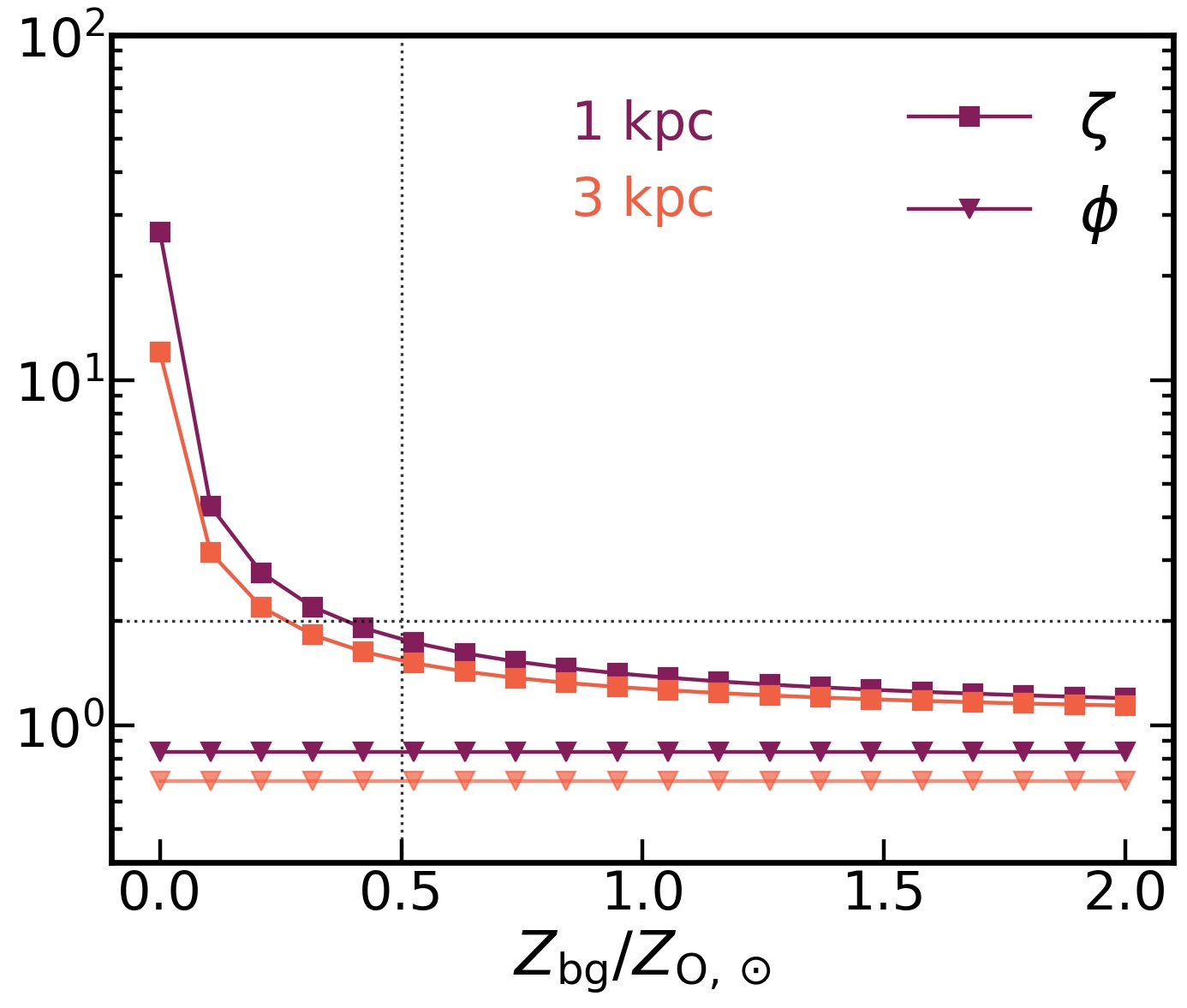}
    \caption{Same as \autoref{fig:zeta_phi_height}, except now showing $\zeta$ and $\phi$ at fixed heights of $1$ and $3$ kpc but for varying $Z_{\rm bg}$. The horizontal dotted line is at $\zeta=2$ indicates where direct SN ejecta and entrained ISM contribute equally to metal outflows; this occurs at $Z_\mathrm{bg}/Z_{\rm{O},\odot} \approx 0.5$.  
    }
    \label{fig:zeta_phi_zbg}
\end{figure}

Once a steady-state has been established, we use the time-averaged outflow properties to estimate the bulk metal loading factors -- \zta~and \fy~ -- using Equations \ref{eqn:zeta} and \ref{eqn:phi}, respectively. We compute both quantities as a function of height, and all the values we discuss are averaged over the time interval $100-116$ Myr, after the outflow mass and metal fluxes have settled to steady-state. We report these values for every run in \autoref{tab:params}. We note here that we use ``net'' outflow fluxes, that include both outflowing and inflowing material, but that the results do not change substantially if we use outflowing material only.

\subsubsection{Metal loading as a function of height and background metallicity}

\zta\ measures the relative enrichment between the outflowing gas and the gas which has been entrained from the ISM in the outflows. In \autoref{fig:zeta_phi_height}, we show \zta\ and \fy\ as functions of height from the mid-plane for FG$2$. We use net outflow rates averaged over $20$ pc slabs around each value of $z$. For the case $Z_\mathrm{bg} = 0$, we find $\zeta\gg1$ indicating that metal outflows are dominated by the highly metal-enriched SN ejecta. \zta\ decreases at large distances from the mid-plane mostly because of increasing metallicity of the entrained gas, since outflow rates do not change significantly with height (see \autoref{fig:outflow_rates}). 

In contrast to the metal loading factor $\zeta$, the yield reduction factor, \fy, quantifies the proportion of metals added by the SN feedback that are immediately lost to outflows. That \fy\ remains high even at large heights (barring the decrease towards the edge of the box which we believe is a result of stochasticity in the simulations) suggests that most of the SN ejected metals might escape the disc and eventually contaminate the CGM.

Because the metal outflow rate and the average metallicity also depend on the level of background enrichment, we expect \zta\ to change with $Z_{\rm bg}$. In \autoref{fig:zeta_phi_zbg} we show the variation of both \zta\ and \fy\ with $Z_{\rm bg}$ at two different heights. As expected, \zta\ decreases with increasing background metallicity because as $Z_\mathrm{bg}$ increases entrained ISM contributes an increasingly large fraction of the outflowing metal flux. At $Z_{\rm bg}=Z_{\rm O,\odot}$, we find $\zeta \approx 1.3$, which corresponds to the metal outflows containing a slightly sub-dominant contribution from direct, unmixed SN ejecta and a stronger contribution from metals entrained from the background ISM. Examining the dependence of $\zeta$ on $Z_\mathrm{bg}$ more broadly, we find that, for outflows typical of Solar Neighbourhood conditions, entrained ISM and direct SN ejecta contribute approximately equally for a background metallicity $Z_\mathrm{bg} \approx 0.5Z_{\rm O,\odot}$  with direct ejecta dominating at lower metallicity and entrained metals at higher metallicity.

We expect that the yield factor should \textit{not} depend on the ISM metallicity $Z_\mathrm{bg}$, and \autoref{fig:zeta_phi_zbg} confirms this expectation: \fy\ is nearly independent of $Z_\mathrm{bg}$, and, as \autoref{fig:zeta_phi_height}, is nearly constant with height as well. A critical conclusion to draw from \autoref{fig:zeta_phi_zbg} is that $\phi$ is quite close to unity, meaning that a significant majority of SN-injected metals are lost to outflows rather than mixing with the ISM.

\subsubsection{Testing Convergence}

\begin{figure}
\begin{center}
    
	\includegraphics[width=\columnwidth]{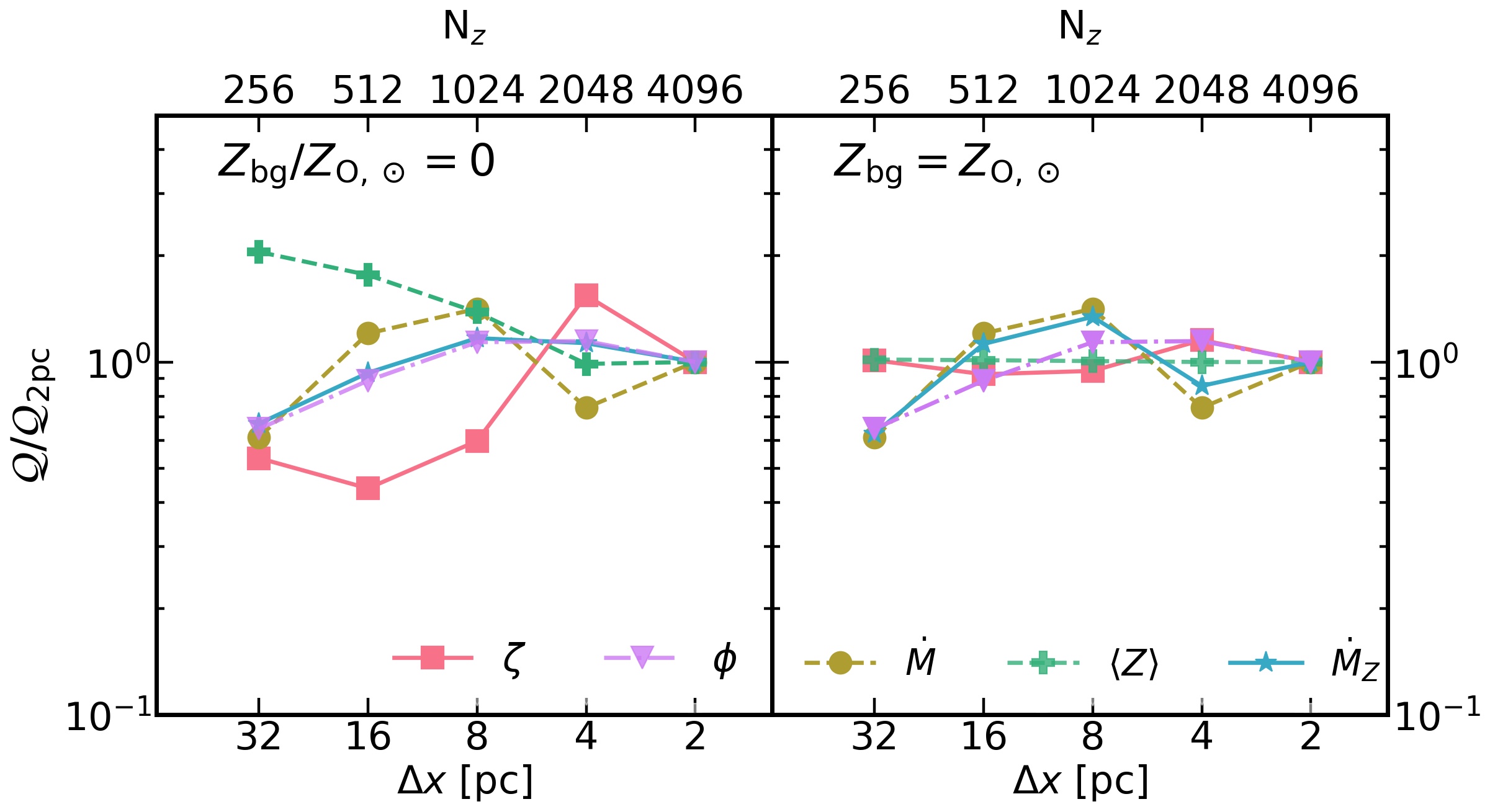}
    \caption{
    Time-averaged values of metal-loading factors $\zeta$ and \fy, mass outflow rate $\dot{M}$, metal outflow rate $\dot{M}_Z$, and mean metallicity $\langle Z\rangle$, all computed for a $20$ pc slab around $|z|=1$ kpc and for $Z_\mathrm{bg} = 0$ (left) and $Z_\mathrm{bg} = Z_{\odot}$ (right), as a function of simulation resolution. The values of all quantities are normalised to the results of run FG$2$, the highest-resolution run. Note that \zta\ is converged even at 32 pc resolution in the case of an enriched background, but does not converge until much higher resolution for the case $Z_\mathrm{bg} = 0$.  
    }
    \label{fig:quant_vs_resolution}
    \end{center}
\end{figure}

Thus far we have focused on results from run FG$2$, our highest resolution run. However, we have not yet established that our results are converged at this resolution, and we have yet to establish convergence criteria. We do so by using the metal loading factor \zta; since this in turn depends on the total mass outflow rate, the metal outflow rate, and the mean metallicity, this implies that we aim for convergence in all these quantities. We conduct a series of runs by progressively increasing the base resolution of the grid and establishing steady state in the outflow rates. We expect to reach convergence eventually because we implement feedback as pure thermal energy without a subgrid recipe. 
Because the quantities related to metal distribution, i.e., $\langle {Z} \rangle$ and $\dot{M_{Z}}$, depend on mixing between hot and warm phases, once the interfaces between two phases are resolved, we should achieve convergence.



\autoref{fig:quant_vs_resolution} shows the variation of the metal loading factor and the yield reduction factor with resolution; we also report the numerical values shown in the figure in \autoref{tab:params}. Apart from these factors, we also show how the mass and metal outflow rates and the average metallicity change with resolution. All quantities shown are temporal averages of the spatial averages described in \autoref{sec:quantify_met_loading} across a $20$ pc slab around the height of $1$ kpc, though other heights yield qualitatively similar results. Because \mzout\ and $\langle Z \rangle$ depend on background metallicity, we show the variation of these quantities for both pristine (left) and Solar enriched (right) backgrounds. For $Z_{\rm bg}=0$, between the lowest and highest resolutions, the mass (metal) outflow rates increase (decrease) by a factor of $\lesssim 2$ as we go from $32$ pc to $2$ pc resolution ($N_z = 256$ to $4096$ cells in the $z$ direction). In the same interval, the average metallicity suffers a steeper decline, resulting in a factor of $\sim 5$ increase in \zta. This points to the importance of resolving the interfaces between the temperature phases where metal exchange primarily occurs -- the mean ISM metallicity $\langle Z\rangle$ is lower in the $Z_\mathrm{bg} = 0$ runs at higher resolution because increasing resolution leads to less numerical mixing between the hot and cold phases, and thus to less metal contamination of the cold gas. 
The left panel of \autoref{fig:quant_vs_resolution} shows that though convergence may been achieved in the mass outflow rate at relatively modest resolution, the metal loading factor may not necessarily be converged as a result of this effect. Consequently, $\zeta$ does not appear to converge until $\approx 4$ pc resolution. By contrast the yield reduction factor, $\phi$, depends only on the metal outflow rate for $Z_{\rm bg}=0$ for which $\zeta\gg 1$. Therefore, its convergence curve follows that of the metal outflow rate.

A high background metallicity erases almost all variation in \zta, \mzout, and $\langle Z \rangle$ with resolution, such that it appears that these quantities are converged even at $32$ pc resolution. There is but slight variation in \mzout\ of factor $\lesssim 0.5$ which translates into similar variation in \fy. We stress here that for an enriched background, it is easier to achieve convergence in metal outflow rates and consequently the metal loading and the yield reduction factors, simply because at higher background metallicity numerical diffusion from the hot phase into the cool ISM represents a smaller perturbation.

\subsection{Phase distribution of gas and metals}

\begin{figure*}
	\includegraphics[width=\textwidth]{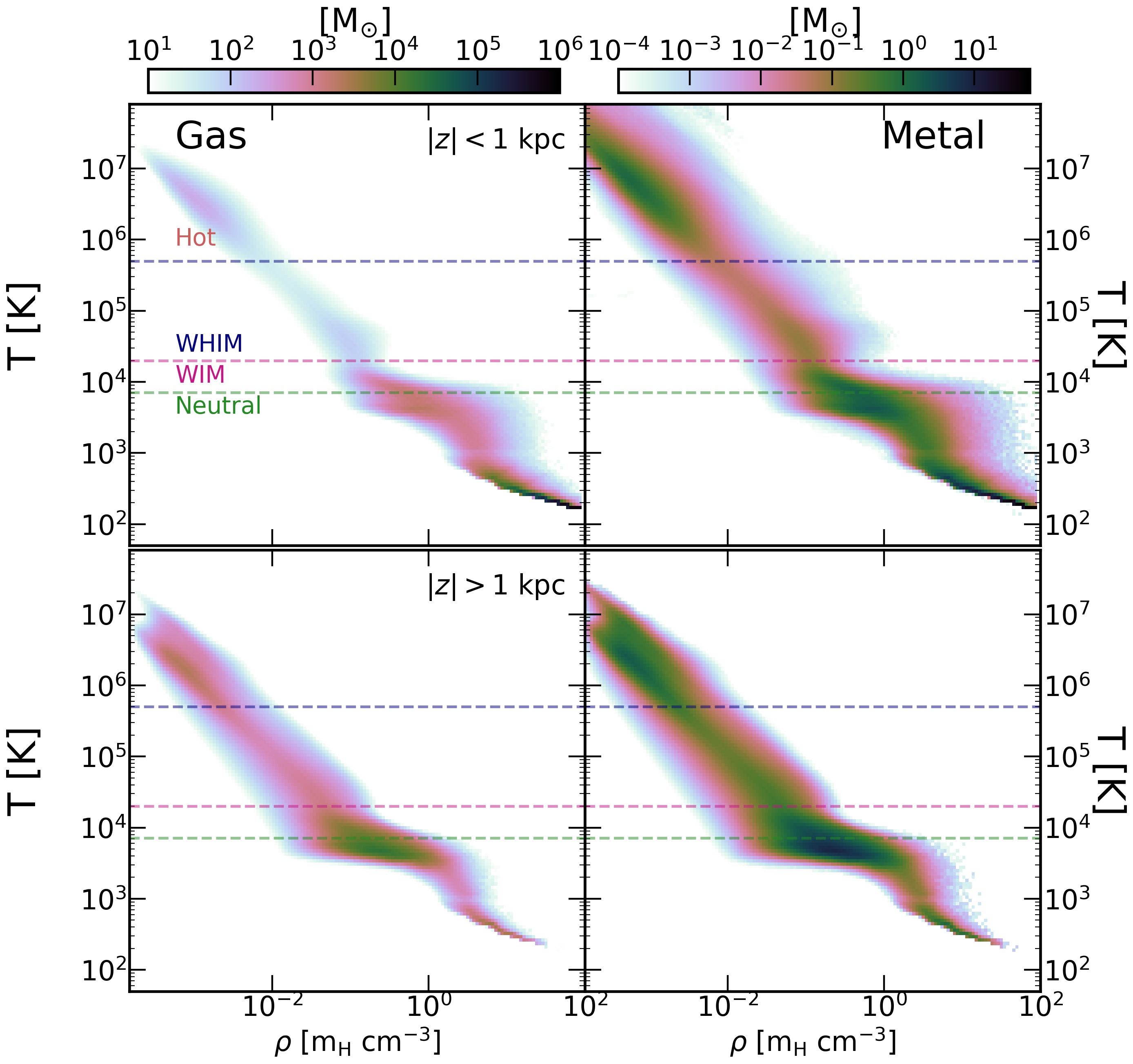}
    \caption{Mass-weighted, time-averaged temperature-density histograms for gas (left) and metals (right), averaged over the regions $|z|<1$ kpc (top) and $|z|>1$ kpc (bottom); metal masses are computed for the case $Z_{\rm bg}=0$. The horizontal lines indicate the temperature thresholds for separating the neutral phase (which combines the CNM, UNM and WNM), the WIM, WHIM, and the hot phase.  In regions closer to the midplane, cold and dense phase dominates the mass content, while for the outflowing gas mass shifts to higher temperatures.
    }
    \label{fig:temp-dens-histo}
\end{figure*}

\begin{figure}
	\includegraphics[width=\columnwidth]{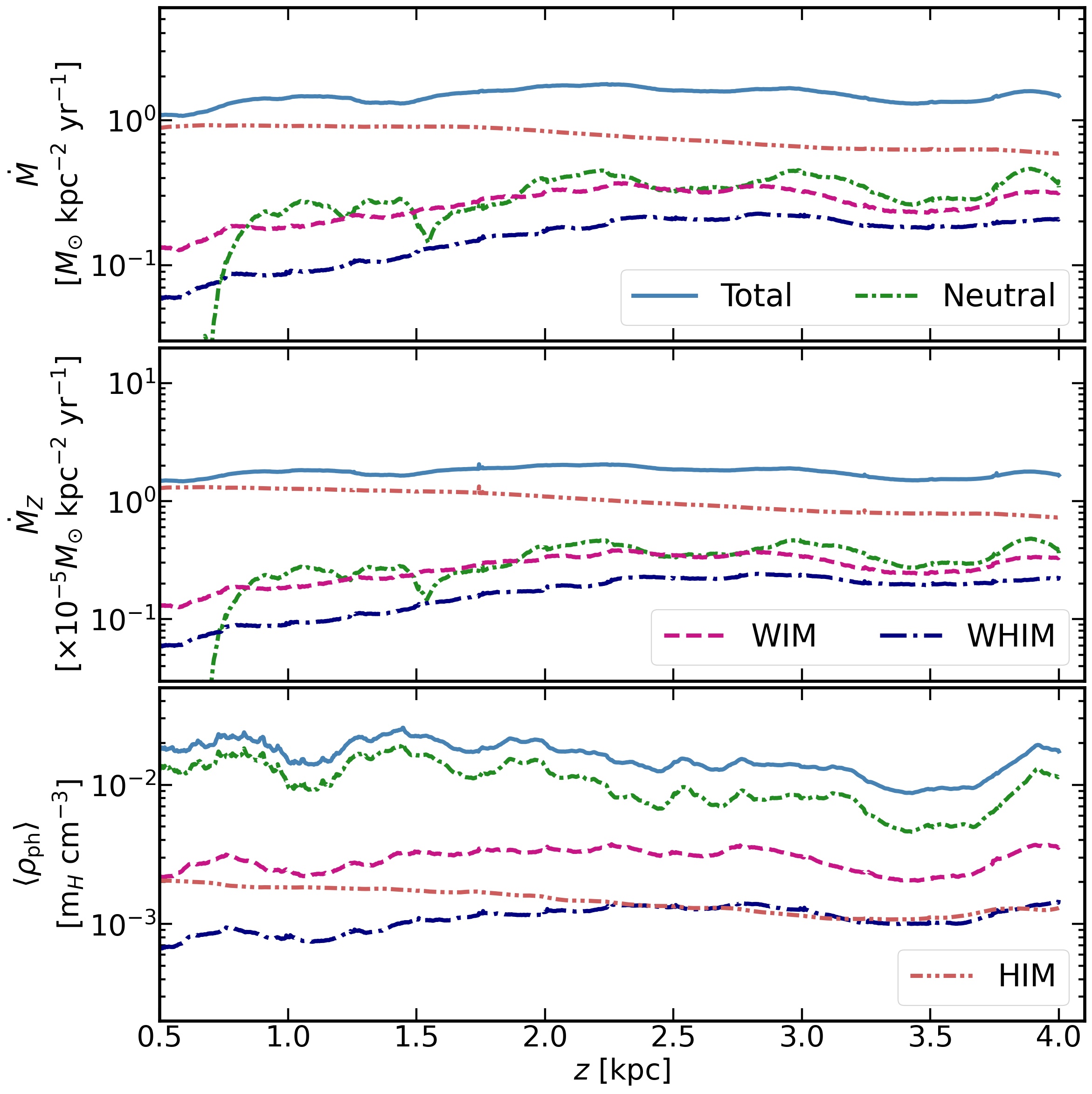}
    \caption{The phase separated fluxes of mass (top) and metal (middle) and the average mass density (bottom). Though HIM is responsible for most of the mass flux close to the disc, the balance shifts towards the WIM and WNM phases at larger $z$ values. In case of the metal flux, the hot material dominates all the way till the edge of the simulation domain. 
    }
    \label{fig:flux_by_phase}
\end{figure}

By partitioning the gas into different temperature bins we assess the contribution of the different phases towards mass and metal outflow fluxes. \autoref{fig:temp-dens-histo} shows the distribution of gas in the density-temperature plane, weighted by both total mass (top) and metal mass (bottom) and both close to the disc ($|z|<1$ kpc, left) and far from it ($|z|>1$ kpc, right); as usual, we show the case $Z_\mathrm{bg} = 0$. Gas and metal mass both clearly populate at least six different phases with different temperatures $T$, which are imposed by our cooling curve. As discussed in \citet{Wibking23a}, these phases are:
\begin{enumerate}
    \item Cold neutral medium (CNM) - $T/\mbox{K}<980$
    \item Unstable neutral medium (UNM) - $980< T/\mbox{K}\leq 4126$
    \item Warm neutral medium (WNM) - $4126 < T/\mbox{K} \leq 7105$
    \item Warm ionized medium (WIM) - $7105<T/\mbox{K}\leq 2\times 10^4$
    \item Warm-hot ionized medium (WHIM) - $2\times 10^4<T/\mbox{K} \leq 5\times 10^5$
    \item Hot ionised medium (HIM) - $T/\mbox{K}>5\times 10^5$.
\end{enumerate}
Examining \autoref{fig:temp-dens-histo}, we see that all these phases are populated both close to the disc and far from it, but that the relative contributions vary with height.
Both gas and metal masses in the region closer to the disc are dominated by CNM and WNM, while the CNM is much sparser in the extraplanar regions. In the $|z|>1$ kpc region, the bulk of the mass lies in the WNM and WIM, though this region also hosts more WHIM and HIM than the region closer to the disc. We note here that within $z<|1|$ kpc HIM amd WHIM host more metals relative to the gas mass they carry. In the extra-planar regions, HIM and WHIM host most of the metals. For HIM this shows that most of the metals do not mix with the ISM and quickly escape into extra-planar region. WHIM acquires metals by cooling of HIM and heating of cooler phases by means of mixing. In the remainder of this section we examine the properties of the outflow as a function of phase; because CNM and UNM are a subdominant (but, we emphasise, not completely negligible) component in the outflow region, for simplicity in the remainder of this section we will group these phases together with WNM as a single neutral phase.

\subsubsection{Phase distribution by mass and flux}

While \autoref{fig:temp-dens-histo} shows the distribution by \textit{mass}, it is interesting to contrast this with the distribution by \textit{flux}. To explore this, we show the distribution of mass (top) and metal (middle) fluxes passing through $2$ pc thick slabs at different heights in \autoref{fig:flux_by_phase}. For comparison, in the bottom panel of \autoref{fig:flux_by_phase} we show the partial density in each phase, defined as the total mass of material in that phase divided by the volume containing all material (as opposed to the mass of each phase divided only by the volume it occupies).

The phase structures of mass and metal fluxes are quite different. Focusing first on the former, the neutral, WIM and WHIM phases carry $\sim 10\%$ of mass flux at $1$ kpc but this increases to $50\%$ at $4$ kpc; by contrast, that the neutral phase dominates the mass budget at all heights. Given that the cooler phases are responsible for most of the mass at all heights, and that the increase in their partial densities with height is much smaller than the increase in their contribution to the flux, we can conclude that these phases are being accelerated, rather than forming via condensation of hot gas. We note here under ballistic conditions we should expect a decrease in momentum of the cooler phases, neutral in particular, simply due to the inertia it carries. In fact, we find the opposite, a point whose significance we discuss in \autoref{sec:discussion}.

With regard to metals, the key conclusion to be drawn from \autoref{fig:flux_by_phase} is that most of the metal flux is carried in the hot phase of the outflows, though the proportion changes with height. Closer to the midplane, at $1$ kpc, hot outflows carry nearly $90\%$ of the metal flux, while at the larger height this decreases but remains $>50\%$ of the total. This is in marked contrast to the distribution of metals by mass shown in \autoref{fig:temp-dens-histo}, where even at $|z| > 1$ kpc most of the metal \textit{mass} lies in gas with $T \lesssim 10^4$ K. This difference can be attributed to the velocity structure of the gas: while the hot phase contains less metal mass, its outward velocity is much larger, and thus it carries a majority of the metal flux. The situation is quite different for the mass flux, which is predominantly hot at $|z| = 1$ kpc, but where the balance shifts in favour of warm (T$\lesssim 10^4$ K) as gas moves from $1$ kpc to $4$ kpc.

\subsubsection{Phase-wise metallicity}\label{sec:phase_wise_metallicity}

\begin{figure*}
	\includegraphics[width=\textwidth]{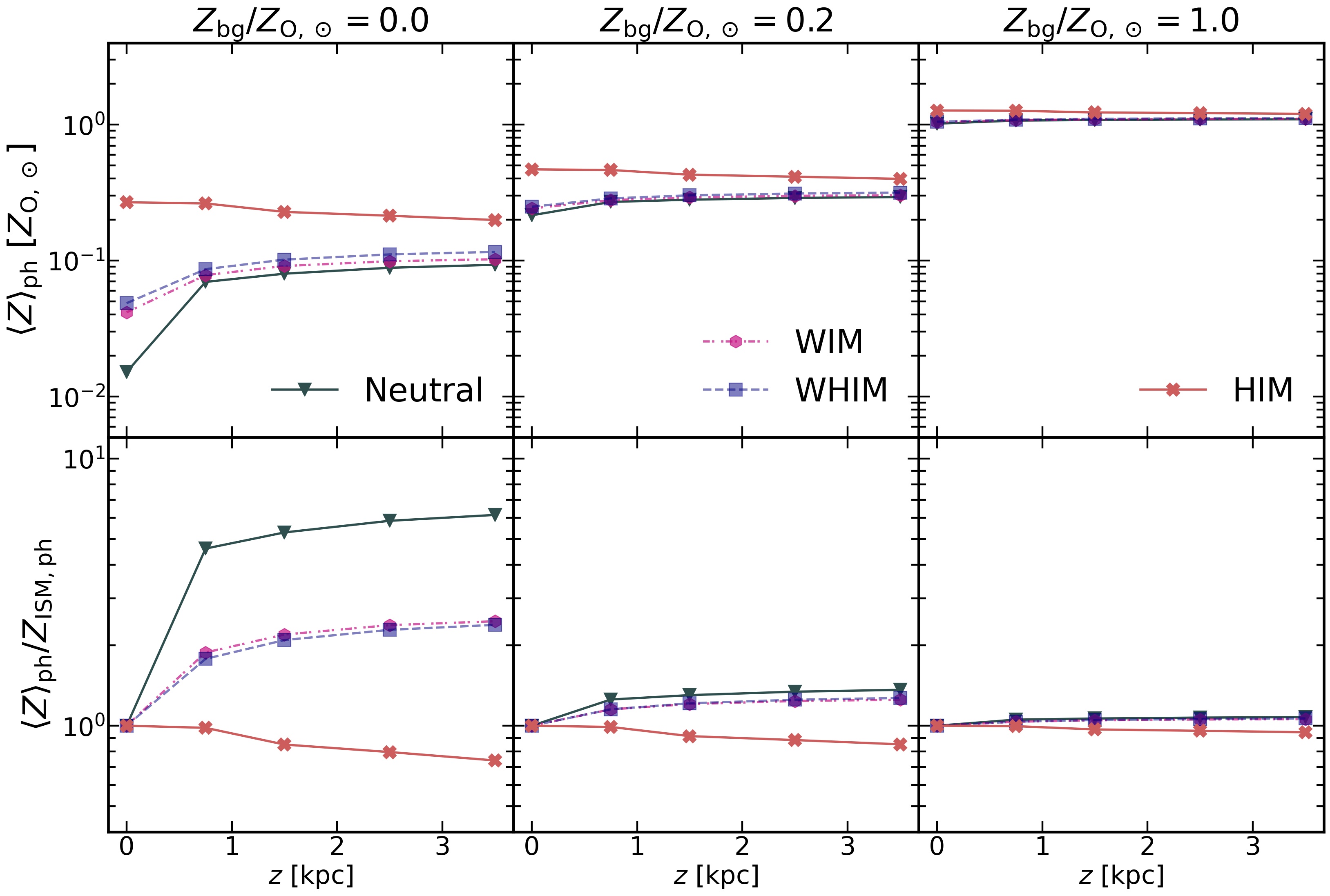}
    \caption{Mass-weighted metallicities (\autoref{eqn:avg_z}) in $1$ kpc-wide sections of the simulation domain, separated by temperature phases. The top row shows metallicity on an absolute scale, while the bottom shows metallicity normalised to the mean metallicity of each phase at $z=0$.
    }
    \label{fig:phase-metallicity}
\end{figure*}

Thus far we have examined the distribution of mass and flux with respect to phase; however, these quantities are not easily accessible via observations. Instead, what observations can probe is differences in the abundance of one phase of the outflowing gas versus another, and differences with respect to height within a single phase. We investigate these differences by computing time-averaged abundances (\autoref{eqn:avg_z}) over $1$ kpc-wide regions of the simulation domain at times after $100$ Myr, once steady-state has been established. We compute these averages separately for each of the phases listed above, though for convenience and to avoid cluttering our plots we again group all the neutral phases (CNM, UNM, WNM) together for this purpose.

We plot the average metallicities as a function of distance from the mid-plane for each phase in the top row of \autoref{fig:phase-metallicity}; the three columns show three different background 
metallicities, $Z_\mathrm{bg}/Z_\mathrm{O,\odot} = 0, 0.2$ (comparable to the metallicity of the Small Magellanic Cloud and of dwarf starbursts whose outflows have been studied in observations), and $1$. We see that the hot phase, carrying the fresh SN ejecta, is the most metal-rich at all heights, but that the difference between its metallicity and the mean metallicity of the cooler phases is a function of both height and background metallicity; close to the mid-plane and for $Z_\mathrm{bg} = 0$, the hot phase is as much as $\approx 5\times$ more metal-rich than any of the cooler phases, while for $Z_\mathrm{bg} = Z_\mathrm{O,\odot}$, the difference drops to at most tens of percent. At an intermediate metallicity of $Z_\mathrm{bg} = 0.2 Z_\mathrm{O,\odot}$, the difference is a factor of $\sim 2$ at low heights, dropping to tens of percent by $\approx 3.5$ kpc. In general the fact that we see both HIM metallicity decreasing and cool phase metallicity increasing with height indicates that there must be material exchanged between the phases in both directions. That is, some metal-poor cool gas must be evaporating into the hot phase in order to explain why the mean HIM metallicity decreases, while some hot material must be condensing into the cooler phases to explain why the WIM, WHIM, and neutral metallicity increases. In the case $Z_\mathrm{bg} = Z_\mathrm{O,\odot}$ where the neutral and hot phases are at similar mean metallicity even at $z=0$, this exchange has little effect on either phase, while its effects are much more dramatic for the case $Z_\mathrm{bg} = 0$, where the hot and neutral phases have very different metallicities near the mid-plane. However, recalling the discussion of mass and metal fluxes in the previous section, we remind readers that, while this exchange occurs, it is not enough to significantly alter which phases carry the bulk of the metal flux.

While the difference in metallicity between phases is of obvious interest from the standpoint of physical interpretation, it is also difficult to probe with observations due to the challenge of cross-calibrating absolute metallicities between, e.g., X-ray and optical data. For this reason, an alternative quantity that is often reported instead is the metallicity of the outflow relative to the ISM, or the variation of metallicity with distance from the galactic plane, within a single phase. The bottom panel of \autoref{fig:phase-metallicity} identical to the top but instead of showing the absolute metallicity, we show metallicity normalised to the mid-plane value. Predictably, for pristine backgrounds, the neutral phase experiences the highest enrichment with respect to its mid-plane metallicity, while the hot phase becomes more metal-poor, albeit by a smaller amount. The relative change in the metallicity of the cool phases is $\approx 50\%$ for $Z_\mathrm{bg} = 0.2Z_\mathrm{O,\odot}$, and this drops to $\lesssim 10\%$ for $Z_\mathrm{bg} = Z_\mathrm{O,\odot}$. The latter is likely unmeasurable at the accuracy of current metallicity diagnostics, but the former may well be observable, and indeed may already have been observed, a point to which we return below.

\section{Discussion}
\label{sec:discussion}

\subsection{Implications for the origin of the mass-metallicity and mass-metallicity gradient relations}

The strongest result from our work is \autoref{fig:zeta_phi_zbg} showing that the yield reduction factor $\phi\approx 0.8$ for star formation in Solar neighborhood-like conditions, i.e., $\approx 80\%$ of the metals in SN ejecta are never mixed with the surrounding ISM and instead escape with the outflowing gas. An important subsidiary point is that this is true despite the fact that the metal loading factor $\zeta$ that characterises the ratio of outflow metallicity to ISM metallicity is relatively modest for Solar metallicity backgrounds. That is, the fact that $\zeta \lesssim 2$ for a metal-rich galaxy does \textit{not} mean that metal loading is modest, as some authors have argued \citep{Kim+20}. Instead, high $\phi$ coexists with moderate $\zeta$ simply because for a very metal-rich background and moderately mass-loaded winds, even very high SN metal loss only enhances wind metallicity mildly compared to ISM metallicity.

Quick expulsion of SN-produced metals is a key piece of physics for understanding the mass-metallicity relation (MZR) and the mass-metallicity gradient relation (MZGR), particularly in dwarf galaxies \citep{Sharda21a, Sharda21b}. Analytical models for these relations find that the observed relatively flat MZGR at low galaxy masses can be understood only if $\phi$ is near unity, as we find. Other semi-analytical models and analysis of cosmological simulations have drawn similar conclusions about metal retention in the ISM of dwarfs \citep{Ma+16, Pandya+21}. Our work strongly supports the view that the paucity of metals in the ISM of dwarf galaxies is a direct result of heavy metal loading of outflows coupled with moderate mass loading \citep[e.g.,][]{Forbes19a}, rather than extreme mass loading as has been suggested elsewhere \citep{Dave+13}.

Preferential ejection of SN-produced metals has also been invoked to explain how the CGM of star-forming galaxies came to hold nearly as much oxygen as the disc \citep{Tumlinson+11, Peeples+14}. \cite{Tumlinson+11} posits that in order to enrich the CGM to the observed levels over a reasonable time scale, most, if not all, of the oxygen produced in SN should be carried out by the outflows, though much of this material is re-accreted over $\sim\mathrm{Gyr}$ timescales and participates in future star formation episodes. This picture is consistent with the narrative set by \autoref{fig:flux_by_phase} that metal flux balance continuously shifts towards warmer phases which may not possess the momentum required to eventually escape the galaxy's potential.

\subsection{Implications for the phase structure of outflows}

A second important conclusion to draw from our work is with regard to the phase structure of outflows. The outflow we produce in our simulations has a structure similar to that proposed by \citet{Thompson16a} and \citet{Krumholz17e}, whereby the cooler phases exist throughout the outflow and are primarily the result of acceleration of pre-existing cool gas out of the plane, rather than re-condensation of hot gas off the plane \citep[e.g.][]{Thompson16b, Schneider18b}. The survival of these cool clouds in the outflow in our simulations appears to be a result of efficient radiative cooling, similar to the effect seen in other simulations with similarly-high spatial resolution \citep[e.g.,][]{Schneider20a, Kim+20}. Such a picture is consistent with recent high spatial resolution observations of outflows, which show that the  neutral and molecular phases are present even at the outflow base \citep[e.g.][]{Leroy15b, Martini18a, Noon23a}, and kinematics suggesting that neutral material accelerates with distance from the galaxy \citep{Yuan23a}. As a result of this structure, neutral or cool ionised material dominates the outflow mass at all heights, but because it accelerates slowly it only becomes a major constituent of the mass flux at heights $\gtrsim 2$ kpc.

Metals add an important dimension to this picture because, especially in the case $Z_\mathrm{bg} = 0$ where the metals are initially present only in the hottest phase, they effectively act as Lagrangian tracers of exchange between phases. The story told by these Lagrangian tracers is that exchange between the HIM phase and the cooler phases is not zero, but is surprisingly small. In particular, recall from \autoref{fig:flux_by_phase} that at no height does the HIM contribute more than $\sim 10\%$ of the mass. Thus even if only $\sim 10\%$ of the essentially metal-free cooler gas were to evaporate into the hot phase, this would be sufficient to dilute its metallicity down by a factor of two. In fact, \autoref{fig:phase-metallicity} shows that the decrease in mean metallicity of the hot phase is considerably smaller than this, meaning that only a few percent of the initially-cooler material can be added to the hot phase over the 4 kpc distance that we track the outflow. Conversely, the fact that the metal flux in the hot phase decreases by a factor $\lesssim 2$ between 0.5 and 4 kpc (cf.~\autoref{fig:flux_by_phase}) implies relatively small loss of hot, metal-rich gas into the cooler phases. Thus the basic picture toward which we are driven is one where, at least out to 4 kpc, the different gas phases for the most part maintain their identities. There is substantial exchange of momentum, as is required to accelerate the cool gas, but not a great deal of exchange of material.

Of course we emphasise that these conclusions apply only up to 4 kpc. A number of simulations with lower resolution but larger volume \citep[e.g.,][]{Schneider20a}, as well as analytic models \citep[e.g.,][]{Fielding22a}, suggest that there should be more exchange between phases at larger heights. We cannot rule out this possibility. However, we also caution that these conclusions are based on simulations and models that do not allow cooling past $10^4$ K, and thus the cool material that becomes hot in these models is assumed to all be in the form of WIM. In fact, we find that even at a height of 4 kpc neutral material represents an equal contribution to the mass flux, and a dominant contribution to the total mass; as noted above, observations of M82 support this conclusion. It is therefore unclear to what extent these models are applicable.

\subsection{Comparison with other theoretical works}

Qualitatively our results reinforce some of the general conclusions of earlier theoretical works, although the physics implemented in each of these is somewhat different, and though our exploration of the effects of varying the background metallicity allows us to draw somewhat different conclusions. Our simulation setup most closely resembles those of \citet{Li&Bryan2017} and \citet{Kim+20}, and our SN injection recipe is similar to \citeauthor{Li&Bryan2017}'s as well, albeit without the contribution of SN Type Ia.  \citeauthor{Li&Bryan2017}'s metal loading factor reduces to our definition of \fy\ in the case of metal poor backgrounds (see \autoref{eqn:phi} in case of $\zeta\gg1$) and we find that both quantities are similar in value -- although it should be noted that the box size used in \cite{Li&Bryan2017} is smaller than ours in volume.

Compared to \citet{Kim+20}, we lack their self-consistent treatment of star formation, but we reach higher resolution over a larger volume. Overall our estimates of the metal outflow rate (\autoref{fig:outflow_rates}) and the enrichment ratio, $\langle Z \rangle/\langle Z_{\rm ISM}\rangle$ (\autoref{fig:phase-metallicity}), are in qualitative agreement with their ``R8'' model. In particular, we find that the hot phase carries most metals, that the metal loading factor $\zeta$ is modest ($\zeta \lesssim 2$), and  an analogous plot for \autoref{fig:flux_by_phase} for a Solar metallicity background shows that at increasingly higher altitudes the metal flux resembles the mass flux as the metal contribution from the entrained gas increases. However, \citet{Kim+20} do not consider the fraction of SN-injected metals lost to outflows, $\phi$, which is arguably more important from the standpoint of the MZR and MZGR than the metal loading factor $\zeta$. They therefore do not reach the two critical conclusions we reach, namely that most SN-injected metals are lost, and that this loss implies much higher metal loading factors in galaxies with lower mean ISM metallicities.

In addition to previous tall box simulations, we can also compare our results to previous simulations of isolated dwarf galaxies. For such a setup, \citet{Emerick+2018} follow detailed chemical evolution of several different ion species and evolve for long timescales, albeit a much lower resolution than we achieve. Their results also point towards poor metal retention, with nearly $\sim 90\%$ of the metals generated by SN feedback being lost, though some of these lost metals may be re-accreted in future. 

\citet{Schneider20a} also simulate an isolated dwarf, though their simulation follows a starburst galaxy with a much higher star formation rate per unit area than our Solar neighbourhood conditions. Interestingly, they appear to find considerably more rapid phase-mixing than we do. They report that the concentration of the passive scalar that they inject into their hot phase is diluted by more than a factor of two even within $2$ kpc of the galactic plane, whereas we find lesser dilution  even out to $4$ kpc (c.f.~\autoref{fig:phase-metallicity}). The cause of the difference is unclear. One obvious candidate is resolution, since at their resolution of $5$ pc we find that $\zeta$ is still not fully converged, and lower resolution promotes mixing. However, there are other possible explanations as well, including the differences in star formation rate, initial conditions, and problem geometry between the two simulations, and differences in the hydrodynamic scheme, to which mixing can be sensitive -- in particular, their scheme uses PLM reconstruction, which is lower-order than the PPM method we use in \textsc{Quokka}, and thus is likely to produce stronger numerical mixing.

\subsection{Comparison with observations}

Results from our work can be directly compared with observations in both optical/UV and X-ray bands. For the former, \citet{Chisholm+2018} measures the metallicity of outflowing gas in galaxies covering several orders of magnitude in mass. Since they use UV absorption, they are able to trace the phases closest to the WIM and WHIM phases described in \autoref{sec:phase_wise_metallicity}. Their measurements support the conclusion that the outflowing gas is heavily metal loaded with respect to the host's ISM, with the amount of metal loading being larger for metal-poor dwarf galaxies than for more metal-rich galaxies. This is at least qualitatively consistent with our findings, in particular \autoref{fig:phase-metallicity}, where we find that the for WIM and WHIM the metallicity is larger in outflowing gas ($|z|\gtrsim 1$ pc) than in midplane gas, but that the difference decreases as the overall galaxy metallicity increases. Interestingly, their entrainment fraction, by which they estimate the fraction of metals in the outflows arising from entrained ISM gas, is $\gtrsim 0.8$. This at might first seem at odds with our conclusions that, at least in dwarfs, $\zeta \gg 2$, i.e., direct SN ejecta dominate the outflow; \citeauthor{Chisholm+2018}'s result corresponds to $\zeta < 2$. However, the contradiction is resolved if we recall from \autoref{fig:flux_by_phase} that WIM and WHIM together carry roughly half the mass flux by only $\approx 10\%$ of the total metal flux, precisely because most of the direct SN-ejected metals are carried in the hot phase do not mix into the WHIM or WIM. Consequently, our simulation is consistent with \citeauthor{Chisholm+2018}'s conclusion that entrained metals dominate, \textit{provided} that we recognise that this conclusion is limited to the phases that are accessible via UV spectroscopy, and is not true of the outflow as a whole. This finding thus highlights the importance of combining observations that probe more than one phase.

With regard to X-rays, \citet{Lopez+2020} analyse \textit{Chandra} observations of the outflowing gas in M82. They follow the warm-hot and the hot phases up to a distance of $\sim 3$ kpc from the disc, traced by O (along with Ne, Mg, Fe) and Si (and also S), respectively. The  abundances of both these phases is nearly flat at $1-1.5$ times the Solar value outside the central injection radius, identified as $500$ pc. Such a trend similar to the $Z_{\mathrm{bg}}=Z_\mathrm{O,\odot}$ panel in \autoref{fig:phase-metallicity}, both qualitatively and quantitatively, though we caution about putting too much weight on this agreement given that our simulation conditions are intended to represent the Solar neighbourhood, not a starburst such as M82. The X-ray surface brightness maps show a steady decline in X-ray luminosity towards regions of higher altitude. Though we plan to compute the X-ray emission properties of our simulations in detail in a later paper, we can predict a similar trend on the basis of \autoref{fig:slice_plot}, noting that gas is hotter closer to the disk. \citet{Lopez+2020} fit models to the spectra from different regions of the wind extract temperature of the emitting gas. They find that the spectra in most regions outside the disc are consistent with the presence of gas at two distinct temperatures, $\sim 0.4-0.6$ keV and $\sim 0.8-1.7$ keV. As can be seen from \autoref{fig:temp-dens-histo}, we also predict significant amount of extra-planar gas in the HIM, which can be as hot as $\sim 10^7$ K ($=1$ keV).

\subsection{Caveats}

Some pieces of physics that may affect outflow properties and phase structures are not yet implemented in QED, and we therefore pause here to note these caveats. One is self-consistent star-formation and pre-SN feedback. For instance, it has been suggested that clustering of SNe may alter the mass loading of the outflows \citep{Smith+21}, and that this in turn is regulated by pre-SN feedback \citep[e.g.,][]{Jeffreson21b}. Any increase in mass loading will affect the metal loading factors of galaxies with a non-zero background metallicity, and it is conceivable that clustering also directly alters phase mixing. However, we note that this effect is less important in simulations that include pre-SN feedback than in earlier ones omitting it.

A second missing piece of physics is cosmic rays (CRs), which some authors find are capable of driving outflows with factor unity of mass loading \citep{Girichidis+16, Pakmor+16, Simpson+16}. CRs drive cooler and denser outflows implying they affect mostly the cooler phases of the outflows \citep{Girichidis+18}, which will be responsible for metal outflows only in non-zero background metallicity. The true importance of CRs for outflow driving is, however, extremely uncertain due to its dependence on poorly-constrained parameters of CR transport \citep[e.g.,][]{Crocker21a, Crocker21b}.

Finally, all gas in QED cools at a rate identical to gas at Solar metallicity -- that is, our cooling is not computed self-consistently with the spatially-varying metallicity. Gas entrained from a sub-solar metallicity ISM should cool more slowly, which might affect the overall phase structure of the outflows. However, it should be noted that in the case of a sub-solar metallicity background, the contribution from this phase to the overall metal loading is also reduced. In future work, we intend to remove the inconsistency as we explore how \zta\ and \fy\ vary with environment. Upcoming iterations of QED will also include a chemistry network that to follow the evolution of individual species rather than lumping all metals together.

\section{Summary and conclusions}

We present results from 3D high-resolution simulations of a patch of a Milky Way-mass galaxy using AMR-based code \textsc{Quokka}, optimised for GPUs. Thanks to this optimisation, QED simulations are able to reach a combination of resolution, volume, and run duration that exceeds any published to date. Our initial setup comprises an initially uniform gas disc with properties modelled on the Solar neighbourhood. Supernova feedback is injected by adding pure thermal energy to cells at a rate consistent with the expect SN rate for the gas surface density, and at random locations drawn from a Gaussian height distribution. We tag the SN ejecta with a passive scalar, representing SN-produced metals, which is then advected with gas. This enables us to track the eventual fate of the metals injected into the ISM by SNe.

We find that the simulation quickly develops large scale outflows, which settle into approximately steady state mass and metal outflow rates after $\approx 100$ Myr. We quantify the metal loading of the outflow in this steady state in terms of two dimensionless factors, \zta\ and \fy. The former is the classical ``metal loading'' factor and describes the enhancement in outflow metal flux compared to what would be expected if the outflow consisted purely of entrained ISM (i.e., with no contribution from unmixed SN ejecta), while the latter quantifies the fraction of metals ejected by SN that end up in metal outflows without ever mixing with the ISM.

Our main findings are:
\begin{enumerate}
    \item The metal loading factor \zta\ is greater than unity, meaning that the outflowing gas carries more metal flux than would be expected if it consisted solely of entrained ISM. However, the amount by which \zta\ exceed unity depends on the background metallicity of the galaxy. Entrained metals dominate the metal flux ($\zeta < 2$) for galaxies with metallicities $\gtrsim 20\%$ of Solar, while direct SN ejecta dominate in more metal-poor systems.  
    
    \item By contrast the yield reduction factor, \fy, which characterises the fraction of SN-injected metals that are lost promptly to the outflow, does not depend on the background metallicity and is fairly close to unity, $\phi\approx 0.8-0.9$. Thus most of the metals produced by stars leave the disc of the galaxy promptly. Theoretical models for the origin of the mass-metallicity and mass-metallicity gradient relations favour such large values of $\phi$, and our simulation results provide physical backing to the large values indirectly inferred from these models.
    
    \item The phase distributions of mass, mass flux, and metal flux are all very different in metal-poor galaxies. In such galaxies, metals are mostly carried in the fast-moving hot phase, but this phase supplies only $\approx 50\%$ of the mass flux and $\lesssim 10\%$ of the mass; instead, most mass resides in neutral gas. These differences are a result of there being relatively little exchange of mass between the phases. However, as the background metallicity increases, the differences between the phase balance of the mass and metal flux diminishes, because as entrained ISM becomes more metal-rich it constitutes a larger and larger share of the total metal flux.

    \item Variations in the outflow metallicity between different phases at fixed height, and within a single gas phase as a function of height, provide a direct and powerful diagnostic of outflow physics, one for which our simulations make definite predictions. In particular, we find that the hot phase should be more metal-rich than warm ionised or neutral gas at fixed height, and that hot gas metallicity should mildly decrease with height, while warm ionised and neutral gas metallicity increases. All of these effects are magnified at low background metallicity and suppressed at high background metallicity.

    \item Capturing metal loading and the balance of metals between different ISM phases in numerical simulations requires very high resolution. We find that these quantities only converge at resolutions of $\approx 2-4$ pc, while lower-resolution simulations tend to underestimate the extent of metal-loading.
\end{enumerate}
  
\section*{Acknowledgements}

The authors acknowledge function from the Australian Research Council through awards FL220100020 and DP230101064. This research/project was undertaken with the assistance of resources and services from the National Computational Infrastructure (NCI), which is supported by the Australian Government, and from the Pawsey Supercomputer Research Centre, which supported by the Australian Government and the Government of Western Australia.

\section*{Data Availability}

The data underlying this article will be shared on reasonable request to the corresponding author. The \textsc{Quokka} code used for this work is freely available from \url{https://github.com/quokka-astro/quokka}.



\bibliographystyle{mnras}
\bibliography{references} 

\begin{thebibliography}{}
\makeatletter
\relax
\def\mn@urlcharsother{\let\do\@makeother \do\$\do\&\do\#\do\^\do\_\do\%\do\~}
\def\mn@doi{\begingroup\mn@urlcharsother \@ifnextchar [ {\mn@doi@}
  {\mn@doi@[]}}
\def\mn@doi@[#1]#2{\def\@tempa{#1}\ifx\@tempa\@empty \href
  {http://dx.doi.org/#2} {doi:#2}\else \href {http://dx.doi.org/#2} {#1}\fi
  \endgroup}
\def\mn@eprint#1#2{\mn@eprint@#1:#2::\@nil}
\def\mn@eprint@arXiv#1{\href {http://arxiv.org/abs/#1} {{\tt arXiv:#1}}}
\def\mn@eprint@dblp#1{\href {http://dblp.uni-trier.de/rec/bibtex/#1.xml}
  {dblp:#1}}
\def\mn@eprint@#1:#2:#3:#4\@nil{\def\@tempa {#1}\def\@tempb {#2}\def\@tempc
  {#3}\ifx \@tempc \@empty \let \@tempc \@tempb \let \@tempb \@tempa \fi \ifx
  \@tempb \@empty \def\@tempb {arXiv}\fi \@ifundefined
  {mn@eprint@\@tempb}{\@tempb:\@tempc}{\expandafter \expandafter \csname
  mn@eprint@\@tempb\endcsname \expandafter{\@tempc}}}

\bibitem[\protect\citeauthoryear{{Aguirre}, {Hernquist}, {Schaye}, {Weinberg},
  {Katz}  \& {Gardner}}{{Aguirre} et~al.}{2001}]{Aguirre+2001}
{Aguirre} A.,  {Hernquist} L.,  {Schaye} J.,  {Weinberg} D.~H.,  {Katz} N.,
  {Gardner} J.,  2001, \mn@doi [\apj] {10.1086/323070}, \href
  {https://ui.adsabs.harvard.edu/abs/2001ApJ...560..599A} {560, 599}

\bibitem[\protect\citeauthoryear{{Andersson}, {Agertz}, {Renaud}  \&
  {Teyssier}}{{Andersson} et~al.}{2023}]{Andersson23a}
{Andersson} E.~P.,  {Agertz} O.,  {Renaud} F.,   {Teyssier} R.,  2023, \mn@doi
  [\mnras] {10.1093/mnras/stad692}, \href
  {https://ui.adsabs.harvard.edu/abs/2023MNRAS.521.2196A} {521, 2196}

\bibitem[\protect\citeauthoryear{{Asplund}, {Grevesse}, {Sauval}  \&
  {Scott}}{{Asplund} et~al.}{2009}]{Asplund09}
{Asplund} M.,  {Grevesse} N.,  {Sauval} A.~J.,   {Scott} P.,  2009, \mn@doi
  [\araa] {10.1146/annurev.astro.46.060407.145222}, \href
  {http://adsabs.harvard.edu/abs/2009ARA%26A..47..481A} {47, 481}

\bibitem[\protect\citeauthoryear{{Belfiore} et~al.,}{{Belfiore}
  et~al.}{2017}]{Belfiore17a}
{Belfiore} F.,  et~al., 2017, \mn@doi [\mnras] {10.1093/mnras/stw3211}, \href
  {http://adsabs.harvard.edu/abs/2017MNRAS.466.2570B} {466, 2570}

\bibitem[\protect\citeauthoryear{{Cameron} et~al.,}{{Cameron}
  et~al.}{2021}]{Cameron+2021}
{Cameron} A.~J.,  et~al., 2021, \mn@doi [\apjl] {10.3847/2041-8213/ac18ca},
  \href {https://ui.adsabs.harvard.edu/abs/2021ApJ...918L..16C} {918, L16}

\bibitem[\protect\citeauthoryear{{Chabrier}}{{Chabrier}}{2001}]{Chabrier2001}
{Chabrier} G.,  2001, \mn@doi [\apj] {10.1086/321401}, \href
  {https://ui.adsabs.harvard.edu/abs/2001ApJ...554.1274C} {554, 1274}

\bibitem[\protect\citeauthoryear{{Chisholm}, {Tremonti}  \&
  {Leitherer}}{{Chisholm} et~al.}{2018}]{Chisholm+2018}
{Chisholm} J.,  {Tremonti} C.,   {Leitherer} C.,  2018, \mn@doi [\mnras]
  {10.1093/mnras/sty2380}, \href
  {https://ui.adsabs.harvard.edu/abs/2018MNRAS.481.1690C} {481, 1690}

\bibitem[\protect\citeauthoryear{{Christensen}, {Dav{\'e}}, {Brooks}, {Quinn}
  \& {Shen}}{{Christensen} et~al.}{2018}]{Christensen18a}
{Christensen} C.~R.,  {Dav{\'e}} R.,  {Brooks} A.,  {Quinn} T.,   {Shen} S.,
  2018, \mn@doi [\apj] {10.3847/1538-4357/aae374}, \href
  {https://ui.adsabs.harvard.edu/abs/2018ApJ...867..142C} {867, 142}

\bibitem[\protect\citeauthoryear{{Creasey}, {Theuns}  \& {Bower}}{{Creasey}
  et~al.}{2015}]{Creasy+15}
{Creasey} P.,  {Theuns} T.,   {Bower} R.~G.,  2015, \mn@doi [\mnras]
  {10.1093/mnras/stu2233}, \href
  {https://ui.adsabs.harvard.edu/abs/2015MNRAS.446.2125C} {446, 2125}

\bibitem[\protect\citeauthoryear{{Crocker}, {Krumholz}  \&
  {Thompson}}{{Crocker} et~al.}{2021a}]{Crocker21a}
{Crocker} R.~M.,  {Krumholz} M.~R.,   {Thompson} T.~A.,  2021a, \mn@doi
  [\mnras] {10.1093/mnras/stab148}, \href
  {https://ui.adsabs.harvard.edu/abs/2021MNRAS.502.1312C} {502, 1312}

\bibitem[\protect\citeauthoryear{{Crocker}, {Krumholz}  \&
  {Thompson}}{{Crocker} et~al.}{2021b}]{Crocker21b}
{Crocker} R.~M.,  {Krumholz} M.~R.,   {Thompson} T.~A.,  2021b, \mn@doi
  [\mnras] {10.1093/mnras/stab502}, \href
  {https://ui.adsabs.harvard.edu/abs/2021MNRAS.503.2651C} {503, 2651}

\bibitem[\protect\citeauthoryear{{Dav{\'e}}, {Finlator}  \&
  {Oppenheimer}}{{Dav{\'e}} et~al.}{2011}]{Dave+13}
{Dav{\'e}} R.,  {Finlator} K.,   {Oppenheimer} B.~D.,  2011, \mn@doi [\mnras]
  {10.1111/j.1365-2966.2011.19132.x}, \href
  {https://ui.adsabs.harvard.edu/abs/2011MNRAS.416.1354D} {416, 1354}

\bibitem[\protect\citeauthoryear{{Dav{\'e}}, {Finlator}  \&
  {Oppenheimer}}{{Dav{\'e}} et~al.}{2012}]{Dave12a}
{Dav{\'e}} R.,  {Finlator} K.,   {Oppenheimer} B.~D.,  2012, \mn@doi [\mnras]
  {10.1111/j.1365-2966.2011.20148.x}, \href
  {http://adsabs.harvard.edu/abs/2012MNRAS.421...98D} {421, 98}

\bibitem[\protect\citeauthoryear{{Duval} et~al.,}{{Duval}
  et~al.}{2016}]{Duval+2016}
{Duval} F.,  et~al., 2016, \mn@doi [\aap] {10.1051/0004-6361/201526876}, \href
  {https://ui.adsabs.harvard.edu/abs/2016A&A...587A..77D} {587, A77}

\bibitem[\protect\citeauthoryear{{Emerick}, {Bryan}, {Mac Low}, {C{\^o}t{\'e}},
  {Johnston}  \& {O'Shea}}{{Emerick} et~al.}{2018}]{Emerick+2018}
{Emerick} A.,  {Bryan} G.~L.,  {Mac Low} M.-M.,  {C{\^o}t{\'e}} B.,  {Johnston}
  K.~V.,   {O'Shea} B.~W.,  2018, \mn@doi [\apj] {10.3847/1538-4357/aaec7d},
  \href {https://ui.adsabs.harvard.edu/abs/2018ApJ...869...94E} {869, 94}

\bibitem[\protect\citeauthoryear{{Fielding} \& {Bryan}}{{Fielding} \&
  {Bryan}}{2022}]{Fielding22a}
{Fielding} D.~B.,  {Bryan} G.~L.,  2022, \mn@doi [\apj]
  {10.3847/1538-4357/ac2f41}, \href
  {https://ui.adsabs.harvard.edu/abs/2022ApJ...924...82F} {924, 82}

\bibitem[\protect\citeauthoryear{{Finlator} \& {Dav{\'e}}}{{Finlator} \&
  {Dav{\'e}}}{2008}]{Finlator&Dave2008}
{Finlator} K.,  {Dav{\'e}} R.,  2008, \mn@doi [\mnras]
  {10.1111/j.1365-2966.2008.12991.x}, \href
  {https://ui.adsabs.harvard.edu/abs/2008MNRAS.385.2181F} {385, 2181}

\bibitem[\protect\citeauthoryear{{Forbes}, {Krumholz}, {Goldbaum}  \&
  {Dekel}}{{Forbes} et~al.}{2016}]{Forbes+16}
{Forbes} J.~C.,  {Krumholz} M.~R.,  {Goldbaum} N.~J.,   {Dekel} A.,  2016,
  \mn@doi [\nat] {10.1038/nature18292}, \href
  {https://ui.adsabs.harvard.edu/abs/2016Natur.535..523F} {535, 523}

\bibitem[\protect\citeauthoryear{{Forbes}, {Krumholz}  \& {Speagle}}{{Forbes}
  et~al.}{2019}]{Forbes19a}
{Forbes} J.~C.,  {Krumholz} M.~R.,   {Speagle} J.~S.,  2019, \mn@doi [\mnras]
  {10.1093/mnras/stz1473}, \href
  {https://ui.adsabs.harvard.edu/abs/2019MNRAS.487.3581F} {487, 3581}

\bibitem[\protect\citeauthoryear{{Fragile}, {Murray}  \& {Lin}}{{Fragile}
  et~al.}{2004}]{Fragile+2004}
{Fragile} P.~C.,  {Murray} S.~D.,   {Lin} D. N.~C.,  2004, \mn@doi [\apj]
  {10.1086/425494}, \href
  {https://ui.adsabs.harvard.edu/abs/2004ApJ...617.1077F} {617, 1077}

\bibitem[\protect\citeauthoryear{{Gentry}, {Krumholz}, {Madau}  \&
  {Lupi}}{{Gentry} et~al.}{2019}]{Gentry19a}
{Gentry} E.~S.,  {Krumholz} M.~R.,  {Madau} P.,   {Lupi} A.,  2019, \mn@doi
  [\mnras] {10.1093/mnras/sty3319}, \href
  {https://ui.adsabs.harvard.edu/\#abs/2019MNRAS.483.3647G} {483, 3647}

\bibitem[\protect\citeauthoryear{{Girichidis} et~al.,}{{Girichidis}
  et~al.}{2016}]{Girichidis+16}
{Girichidis} P.,  et~al., 2016, \mn@doi [\apjl] {10.3847/2041-8205/816/2/L19},
  \href {https://ui.adsabs.harvard.edu/abs/2016ApJ...816L..19G} {816, L19}

\bibitem[\protect\citeauthoryear{{Girichidis}, {Naab}, {Hanasz}  \&
  {Walch}}{{Girichidis} et~al.}{2018}]{Girichidis+18}
{Girichidis} P.,  {Naab} T.,  {Hanasz} M.,   {Walch} S.,  2018, \mn@doi
  [\mnras] {10.1093/mnras/sty1653}, \href
  {https://ui.adsabs.harvard.edu/abs/2018MNRAS.479.3042G} {479, 3042}

\bibitem[\protect\citeauthoryear{{Jeffreson}, {Krumholz}, {Fujimoto},
  {Armillotta}, {Keller}, {Chevance}  \& {Kruijssen}}{{Jeffreson}
  et~al.}{2021}]{Jeffreson21b}
{Jeffreson} S. M.~R.,  {Krumholz} M.~R.,  {Fujimoto} Y.,  {Armillotta} L.,
  {Keller} B.~W.,  {Chevance} M.,   {Kruijssen} J.~M.~D.,  2021, \mn@doi
  [\mnras] {10.1093/mnras/stab1536}, \href
  {https://ui.adsabs.harvard.edu/abs/2021MNRAS.505.3470J} {505, 3470}

\bibitem[\protect\citeauthoryear{{Kim} \& {Ostriker}}{{Kim} \&
  {Ostriker}}{2017}]{Kim&Ostriker2017}
{Kim} C.-G.,  {Ostriker} E.~C.,  2017, \mn@doi [\apj]
  {10.3847/1538-4357/aa8599}, \href
  {https://ui.adsabs.harvard.edu/abs/2017ApJ...846..133K} {846, 133}

\bibitem[\protect\citeauthoryear{{Kim} et~al.,}{{Kim} et~al.}{2020}]{Kim+20}
{Kim} C.-G.,  et~al., 2020, \mn@doi [\apj] {10.3847/1538-4357/aba962}, \href
  {https://ui.adsabs.harvard.edu/abs/2020ApJ...900...61K} {900, 61}

\bibitem[\protect\citeauthoryear{{Konami}, {Matsushita}, {Tsuru}, {Gandhi}  \&
  {Tamagawa}}{{Konami} et~al.}{2011}]{Konami+2011}
{Konami} S.,  {Matsushita} K.,  {Tsuru} T.~G.,  {Gandhi} P.,   {Tamagawa} T.,
  2011, \mn@doi [\pasj] {10.1093/pasj/63.sp3.S913}, \href
  {https://ui.adsabs.harvard.edu/abs/2011PASJ...63S.913K} {63, S913}

\bibitem[\protect\citeauthoryear{{Krumholz}, {Thompson}, {Ostriker}  \&
  {Martin}}{{Krumholz} et~al.}{2017}]{Krumholz17e}
{Krumholz} M.~R.,  {Thompson} T.~A.,  {Ostriker} E.~C.,   {Martin} C.~L.,
  2017, \mn@doi [\mnras] {10.1093/mnras/stx1882}, \href
  {http://adsabs.harvard.edu/abs/2017MNRAS.471.4061K} {471, 4061}

\bibitem[\protect\citeauthoryear{{Kuijken} \& {Gilmore}}{{Kuijken} \&
  {Gilmore}}{1989}]{Kuijken89a}
{Kuijken} K.,  {Gilmore} G.,  1989, \mn@doi [\mnras] {10.1093/mnras/239.2.605},
  \href {https://ui.adsabs.harvard.edu/abs/1989MNRAS.239..605K} {239, 605}

\bibitem[\protect\citeauthoryear{{Leroy} et~al.,}{{Leroy}
  et~al.}{2015}]{Leroy15b}
{Leroy} A.~K.,  et~al., 2015, \mn@doi [\apj] {10.1088/0004-637X/814/2/83},
  \href {http://adsabs.harvard.edu/abs/2015ApJ...814...83L} {814, 83}

\bibitem[\protect\citeauthoryear{{Li}, {Bryan}  \& {Ostriker}}{{Li}
  et~al.}{2017}]{Li&Bryan2017}
{Li} M.,  {Bryan} G.~L.,   {Ostriker} J.~P.,  2017, \mn@doi [\apj]
  {10.3847/1538-4357/aa7263}, \href
  {https://ui.adsabs.harvard.edu/abs/2017ApJ...841..101L} {841, 101}

\bibitem[\protect\citeauthoryear{{Lilly}, {Carollo}, {Pipino}, {Renzini}  \&
  {Peng}}{{Lilly} et~al.}{2013}]{Lilly13a}
{Lilly} S.~J.,  {Carollo} C.~M.,  {Pipino} A.,  {Renzini} A.,   {Peng} Y.,
  2013, \mn@doi [\apj] {10.1088/0004-637X/772/2/119}, \href
  {http://adsabs.harvard.edu/abs/2013ApJ...772..119L} {772, 119}

\bibitem[\protect\citeauthoryear{{Lopez}, {Mathur}, {Nguyen}, {Thompson}  \&
  {Olivier}}{{Lopez} et~al.}{2020}]{Lopez+2020}
{Lopez} L.~A.,  {Mathur} S.,  {Nguyen} D.~D.,  {Thompson} T.~A.,   {Olivier}
  G.~M.,  2020, \mn@doi [\apj] {10.3847/1538-4357/abc010}, \href
  {https://ui.adsabs.harvard.edu/abs/2020ApJ...904..152L} {904, 152}

\bibitem[\protect\citeauthoryear{{Ma}, {Hopkins}, {Faucher-Gigu{\`e}re},
  {Zolman}, {Muratov}, {Kere{\v{s}}}  \& {Quataert}}{{Ma} et~al.}{2016}]{Ma+16}
{Ma} X.,  {Hopkins} P.~F.,  {Faucher-Gigu{\`e}re} C.-A.,  {Zolman} N.,
  {Muratov} A.~L.,  {Kere{\v{s}}} D.,   {Quataert} E.,  2016, \mn@doi [\mnras]
  {10.1093/mnras/stv2659}, \href
  {https://ui.adsabs.harvard.edu/abs/2016MNRAS.456.2140M} {456, 2140}

\bibitem[\protect\citeauthoryear{{Mac Low} \& {Ferrara}}{{Mac Low} \&
  {Ferrara}}{1999}]{MaclowAndrea99}
{Mac Low} M.-M.,  {Ferrara} A.,  1999, \mn@doi [\apj] {10.1086/306832}, \href
  {https://ui.adsabs.harvard.edu/abs/1999ApJ...513..142M} {513, 142}

\bibitem[\protect\citeauthoryear{{Martin}, {Kobulnicky}  \& {Heckman}}{{Martin}
  et~al.}{2002}]{Martin02a}
{Martin} C.~L.,  {Kobulnicky} H.~A.,   {Heckman} T.~M.,  2002, \mn@doi [\apj]
  {10.1086/341092}, \href {http://adsabs.harvard.edu/abs/2002ApJ...574..663M}
  {574, 663}

\bibitem[\protect\citeauthoryear{{Martini}, {Leroy}, {Mangum}, {Bolatto},
  {Keating}, {Sandstrom}  \& {Walter}}{{Martini} et~al.}{2018}]{Martini18a}
{Martini} P.,  {Leroy} A.~K.,  {Mangum} J.~G.,  {Bolatto} A.,  {Keating} K.~M.,
   {Sandstrom} K.,   {Walter} F.,  2018, \mn@doi [\apj]
  {10.3847/1538-4357/aab08e}, \href
  {https://ui.adsabs.harvard.edu/abs/2018ApJ...856...61M} {856, 61}

\bibitem[\protect\citeauthoryear{{Melioli}, {de Gouveia Dal Pino}  \&
  {Geraissate}}{{Melioli} et~al.}{2013}]{Melioli+2013}
{Melioli} C.,  {de Gouveia Dal Pino} E.~M.,   {Geraissate} F.~G.,  2013,
  \mn@doi [\mnras] {10.1093/mnras/stt126}, \href
  {https://ui.adsabs.harvard.edu/abs/2013MNRAS.430.3235M} {430, 3235}

\bibitem[\protect\citeauthoryear{{Melioli}, {Brighenti}  \&
  {D'Ercole}}{{Melioli} et~al.}{2015}]{Melioli+2015}
{Melioli} C.,  {Brighenti} F.,   {D'Ercole} A.,  2015, \mn@doi [\mnras]
  {10.1093/mnras/stu2008}, \href
  {https://ui.adsabs.harvard.edu/abs/2015MNRAS.446..299M} {446, 299}

\bibitem[\protect\citeauthoryear{{Mingozzi} et~al.,}{{Mingozzi}
  et~al.}{2020}]{Mingozzi20a}
{Mingozzi} M.,  et~al., 2020, \mn@doi [\aap] {10.1051/0004-6361/201937203},
  \href {https://ui.adsabs.harvard.edu/abs/2020A&A...636A..42M} {636, A42}

\bibitem[\protect\citeauthoryear{{Nomoto}, {Kobayashi}  \& {Tominaga}}{{Nomoto}
  et~al.}{2013}]{Nomoto+2013}
{Nomoto} K.,  {Kobayashi} C.,   {Tominaga} N.,  2013, \mn@doi [\araa]
  {10.1146/annurev-astro-082812-140956}, \href
  {https://ui.adsabs.harvard.edu/abs/2013ARA&A..51..457N} {51, 457}

\bibitem[\protect\citeauthoryear{{Noon}, {Krumholz}, {Di Teodoro},
  {McClure-Griffiths}, {Lockman}  \& {Armillotta}}{{Noon}
  et~al.}{2023}]{Noon23a}
{Noon} K.~A.,  {Krumholz} M.~R.,  {Di Teodoro} E.~M.,  {McClure-Griffiths}
  N.~M.,  {Lockman} F.~J.,   {Armillotta} L.,  2023, \mn@doi [\mnras]
  {10.1093/mnras/stad1890}, \href
  {https://ui.adsabs.harvard.edu/abs/2023MNRAS.524.1258N} {524, 1258}

\bibitem[\protect\citeauthoryear{{Pakmor}, {Pfrommer}, {Simpson}  \&
  {Springel}}{{Pakmor} et~al.}{2016}]{Pakmor+16}
{Pakmor} R.,  {Pfrommer} C.,  {Simpson} C.~M.,   {Springel} V.,  2016, \mn@doi
  [\apjl] {10.3847/2041-8205/824/2/L30}, \href
  {https://ui.adsabs.harvard.edu/abs/2016ApJ...824L..30P} {824, L30}

\bibitem[\protect\citeauthoryear{{Pandya} et~al.,}{{Pandya}
  et~al.}{2021}]{Pandya+21}
{Pandya} V.,  et~al., 2021, \mn@doi [\mnras] {10.1093/mnras/stab2714}, \href
  {https://ui.adsabs.harvard.edu/abs/2021MNRAS.508.2979P} {508, 2979}

\bibitem[\protect\citeauthoryear{{Peeples} \& {Shankar}}{{Peeples} \&
  {Shankar}}{2011}]{Peeples&Shankar2011}
{Peeples} M.~S.,  {Shankar} F.,  2011, \mn@doi [\mnras]
  {10.1111/j.1365-2966.2011.19456.x}, \href
  {https://ui.adsabs.harvard.edu/abs/2011MNRAS.417.2962P} {417, 2962}

\bibitem[\protect\citeauthoryear{{Peeples}, {Werk}, {Tumlinson}, {Oppenheimer},
  {Prochaska}, {Katz}  \& {Weinberg}}{{Peeples} et~al.}{2014}]{Peeples+14}
{Peeples} M.~S.,  {Werk} J.~K.,  {Tumlinson} J.,  {Oppenheimer} B.~D.,
  {Prochaska} J.~X.,  {Katz} N.,   {Weinberg} D.~H.,  2014, \mn@doi [\apj]
  {10.1088/0004-637X/786/1/54}, \href
  {https://ui.adsabs.harvard.edu/abs/2014ApJ...786...54P} {786, 54}

\bibitem[\protect\citeauthoryear{{Poetrodjojo} et~al.,}{{Poetrodjojo}
  et~al.}{2021}]{Poetrodjojo21a}
{Poetrodjojo} H.,  et~al., 2021, \mn@doi [\mnras] {10.1093/mnras/stab205},
  \href {https://ui.adsabs.harvard.edu/abs/2021MNRAS.502.3357P} {502, 3357}

\bibitem[\protect\citeauthoryear{{Ranalli}, {Comastri}, {Origlia}  \&
  {Maiolino}}{{Ranalli} et~al.}{2008}]{Ranalli+2008}
{Ranalli} P.,  {Comastri} A.,  {Origlia} L.,   {Maiolino} R.,  2008, \mn@doi
  [\mnras] {10.1111/j.1365-2966.2008.13128.x}, \href
  {https://ui.adsabs.harvard.edu/abs/2008MNRAS.386.1464R} {386, 1464}

\bibitem[\protect\citeauthoryear{{Rey}, {Katz}, {Cameron}, {Devriendt}  \&
  {Slyz}}{{Rey} et~al.}{2023}]{Rey23a}
{Rey} M.~P.,  {Katz} H.~B.,  {Cameron} A.~J.,  {Devriendt} J.,   {Slyz} A.,
  2023, \mn@doi [arXiv e-prints] {10.48550/arXiv.2302.08521}, \href
  {https://ui.adsabs.harvard.edu/abs/2023arXiv230208521R} {p. arXiv:2302.08521}

\bibitem[\protect\citeauthoryear{{Rodr{\'\i}guez-Gonz{\'a}lez}, {Esquivel},
  {Raga}  \& {Col{\'\i}n}}{{Rodr{\'\i}guez-Gonz{\'a}lez}
  et~al.}{2011}]{Rodriguez+2011}
{Rodr{\'\i}guez-Gonz{\'a}lez} A.,  {Esquivel} A.,  {Raga} A.~C.,   {Col{\'\i}n}
  P.,  2011, \mn@doi [\rmxaa] {10.48550/arXiv.1102.0234}, \href
  {https://ui.adsabs.harvard.edu/abs/2011RMxAA..47..113R} {47, 113}

\bibitem[\protect\citeauthoryear{{Schneider}, {Robertson}  \&
  {Thompson}}{{Schneider} et~al.}{2018}]{Schneider18b}
{Schneider} E.~E.,  {Robertson} B.~E.,   {Thompson} T.~A.,  2018, \mn@doi
  [\apj] {10.3847/1538-4357/aacce1}, \href
  {https://ui.adsabs.harvard.edu/abs/2018ApJ...862...56S} {862, 56}

\bibitem[\protect\citeauthoryear{{Schneider}, {Ostriker}, {Robertson}  \&
  {Thompson}}{{Schneider} et~al.}{2020}]{Schneider20a}
{Schneider} E.~E.,  {Ostriker} E.~C.,  {Robertson} B.~E.,   {Thompson} T.~A.,
  2020, \mn@doi [\apj] {10.3847/1538-4357/ab8ae8}, \href
  {https://ui.adsabs.harvard.edu/abs/2020ApJ...895...43S} {895, 43}

\bibitem[\protect\citeauthoryear{{Sharda}, {Krumholz}, {Wisnioski}, {Forbes},
  {Federrath}  \& {Acharyya}}{{Sharda} et~al.}{2021a}]{Sharda21a}
{Sharda} P.,  {Krumholz} M.~R.,  {Wisnioski} E.,  {Forbes} J.~C.,  {Federrath}
  C.,   {Acharyya} A.,  2021a, \mn@doi [\mnras] {10.1093/mnras/stab252}, \href
  {https://ui.adsabs.harvard.edu/abs/2021MNRAS.502.5935S} {502, 5935}

\bibitem[\protect\citeauthoryear{{Sharda}, {Krumholz}, {Wisnioski}, {Acharyya},
  {Federrath}  \& {Forbes}}{{Sharda} et~al.}{2021b}]{Sharda21b}
{Sharda} P.,  {Krumholz} M.~R.,  {Wisnioski} E.,  {Acharyya} A.,  {Federrath}
  C.,   {Forbes} J.~C.,  2021b, \mn@doi [\mnras] {10.1093/mnras/stab868}, \href
  {https://ui.adsabs.harvard.edu/abs/2021MNRAS.504...53S} {504, 53}

\bibitem[\protect\citeauthoryear{{Sharda}, {Ginzburg}, {Krumholz}, {Forbes},
  {Wisnioski}, {Mingozzi}, {Zovaro}  \& {Dekel}}{{Sharda}
  et~al.}{2023}]{Sharda23a}
{Sharda} P.,  {Ginzburg} O.,  {Krumholz} M.~R.,  {Forbes} J.~C.,  {Wisnioski}
  E.,  {Mingozzi} M.,  {Zovaro} H. R.~M.,   {Dekel} A.,  2023, \mn@doi
  [\mnras~in review] {10.48550/arXiv.2303.15853}, \href
  {https://ui.adsabs.harvard.edu/abs/2023arXiv230315853S} {p. arXiv:2303.15853}

\bibitem[\protect\citeauthoryear{{Simpson}, {Pakmor}, {Marinacci}, {Pfrommer},
  {Springel}, {Glover}, {Clark}  \& {Smith}}{{Simpson}
  et~al.}{2016}]{Simpson+16}
{Simpson} C.~M.,  {Pakmor} R.,  {Marinacci} F.,  {Pfrommer} C.,  {Springel} V.,
   {Glover} S. C.~O.,  {Clark} P.~C.,   {Smith} R.~J.,  2016, \mn@doi [\apjl]
  {10.3847/2041-8205/827/2/L29}, \href
  {https://ui.adsabs.harvard.edu/abs/2016ApJ...827L..29S} {827, L29}

\bibitem[\protect\citeauthoryear{{Smith} et~al.,}{{Smith}
  et~al.}{2017}]{Grackle}
{Smith} B.~D.,  et~al., 2017, \mn@doi [\mnras] {10.1093/mnras/stw3291}, \href
  {https://ui.adsabs.harvard.edu/abs/2017MNRAS.466.2217S} {466, 2217}

\bibitem[\protect\citeauthoryear{{Smith}, {Bryan}, {Somerville}, {Hu},
  {Teyssier}, {Burkhart}  \& {Hernquist}}{{Smith} et~al.}{2021}]{Smith+21}
{Smith} M.~C.,  {Bryan} G.~L.,  {Somerville} R.~S.,  {Hu} C.-Y.,  {Teyssier}
  R.,  {Burkhart} B.,   {Hernquist} L.,  2021, \mn@doi [\mnras]
  {10.1093/mnras/stab1896}, \href
  {https://ui.adsabs.harvard.edu/abs/2021MNRAS.506.3882S} {506, 3882}

\bibitem[\protect\citeauthoryear{{Stevens}, {Read}  \&
  {Bravo-Guerrero}}{{Stevens} et~al.}{2003}]{Stevens03a}
{Stevens} I.~R.,  {Read} A.~M.,   {Bravo-Guerrero} J.,  2003, \mn@doi [\mnras]
  {10.1046/j.1365-8711.2003.06894.x}, \href
  {http://adsabs.harvard.edu/abs/2003MNRAS.343L..47S} {343, L47}

\bibitem[\protect\citeauthoryear{{Thompson} \& {Krumholz}}{{Thompson} \&
  {Krumholz}}{2016}]{Thompson16a}
{Thompson} T.~A.,  {Krumholz} M.~R.,  2016, \mn@doi [\mnras]
  {10.1093/mnras/stv2331}, \href
  {http://adsabs.harvard.edu/abs/2016MNRAS.455..334T} {455, 334}

\bibitem[\protect\citeauthoryear{{Thompson}, {Quataert}, {Zhang}  \&
  {Weinberg}}{{Thompson} et~al.}{2016}]{Thompson16b}
{Thompson} T.~A.,  {Quataert} E.,  {Zhang} D.,   {Weinberg} D.~H.,  2016,
  \mn@doi [\mnras] {10.1093/mnras/stv2428}, \href
  {http://adsabs.harvard.edu/abs/2016MNRAS.455.1830T} {455, 1830}

\bibitem[\protect\citeauthoryear{{Tremonti} et~al.,}{{Tremonti}
  et~al.}{2004}]{Tremonti+2004}
{Tremonti} C.~A.,  et~al., 2004, \mn@doi [\apj] {10.1086/423264}, \href
  {https://ui.adsabs.harvard.edu/abs/2004ApJ...613..898T} {613, 898}

\bibitem[\protect\citeauthoryear{{Tumlinson} et~al.,}{{Tumlinson}
  et~al.}{2011}]{Tumlinson+11}
{Tumlinson} J.,  et~al., 2011, \mn@doi [Science] {10.1126/science.1209840},
  \href {https://ui.adsabs.harvard.edu/abs/2011Sci...334..948T} {334, 948}

\bibitem[\protect\citeauthoryear{{Veilleux}, {Maiolino}, {Bolatto}  \&
  {Aalto}}{{Veilleux} et~al.}{2020}]{Veilleux20a}
{Veilleux} S.,  {Maiolino} R.,  {Bolatto} A.~D.,   {Aalto} S.,  2020, \mn@doi
  [\aapr] {10.1007/s00159-019-0121-9}, \href
  {https://ui.adsabs.harvard.edu/abs/2020A&ARv..28....2V} {28, 2}

\bibitem[\protect\citeauthoryear{{Vijayan}, {Kim}, {Armillotta}, {Ostriker}  \&
  {Li}}{{Vijayan} et~al.}{2020}]{Vijayan+20}
{Vijayan} A.,  {Kim} C.-G.,  {Armillotta} L.,  {Ostriker} E.~C.,   {Li} M.,
  2020, \mn@doi [\apj] {10.3847/1538-4357/ab8474}, \href
  {https://ui.adsabs.harvard.edu/abs/2020ApJ...894...12V} {894, 12}

\bibitem[\protect\citeauthoryear{{Wibking} \& {Krumholz}}{{Wibking} \&
  {Krumholz}}{2022}]{QuokkaMethods}
{Wibking} B.~D.,  {Krumholz} M.~R.,  2022, \mn@doi [\mnras]
  {10.1093/mnras/stac439}, \href
  {https://ui.adsabs.harvard.edu/abs/2022MNRAS.512.1430W} {512, 1430}

\bibitem[\protect\citeauthoryear{{Wibking} \& {Krumholz}}{{Wibking} \&
  {Krumholz}}{2023}]{Wibking23a}
{Wibking} B.~D.,  {Krumholz} M.~R.,  2023, \mn@doi [\mnras]
  {10.1093/mnras/stac2648}, \href
  {https://ui.adsabs.harvard.edu/abs/2023MNRAS.521.5972W} {521, 5972}

\bibitem[\protect\citeauthoryear{{Yuan}, {Krumholz}  \& {Martin}}{{Yuan}
  et~al.}{2023}]{Yuan23a}
{Yuan} Y.,  {Krumholz} M.~R.,   {Martin} C.~L.,  2023, \mn@doi [\mnras]
  {10.1093/mnras/stac3241}, \href
  {https://ui.adsabs.harvard.edu/abs/2023MNRAS.518.4084Y} {518, 4084}

\bibitem[\protect\citeauthoryear{{Zahid}, {Torrey}, {Vogelsberger},
  {Hernquist}, {Kewley}  \& {Dav{\'e}}}{{Zahid} et~al.}{2014}]{Zahid14a}
{Zahid} H.~J.,  {Torrey} P.,  {Vogelsberger} M.,  {Hernquist} L.,  {Kewley} L.,
    {Dav{\'e}} R.,  2014, \mn@doi [\apss] {10.1007/s10509-013-1666-0}, \href
  {https://ui.adsabs.harvard.edu/abs/2014Ap&SS.349..873Z} {349, 873}

\makeatother
\end{thebibliography}








\bsp	
\label{lastpage}
\end{document}